\documentclass{aa}

\usepackage{graphicx}
\usepackage[varg]{txfonts}
\usepackage[urlcolor=blue, colorlinks=true, citecolor=blue, linkcolor=blue]{hyperref}
\usepackage{ulem}
\bibpunct{(}{)}{;}{a}{}{,} 
\usepackage{cleveref}
\usepackage{placeins}
\usepackage[super]{nth}
\usepackage{pifont}

\begin{document}  
\newcommand{\KM}[1]{\textcolor{magenta}{{{#1}}}}
\newcommand{\MF}[1]{\textcolor{red}{{{#1}}}}
\newcommand{\DD}[1]{\textcolor{blue}{{{#1}}}}
\newcommand{\xmark}{\ding{55}}

\newcommand{\orcid}[1]{\href{https://orcid.org/#1}{\includegraphics[width=8pt]{Images/orcid.png}}}

   \title{A comparative study of the fundamental metallicity relation}

   \subtitle{The impact of methodology on its observed evolution}

   \author{F.~Pistis\inst{\ref{ncbj}}
          \and A.~Pollo\inst{\ref{ncbj}, \ref{ju}}
          \and M.~Figueira\inst{\ref{ncbj}, \ref{copernicus_torun}}
          \and D.~Vergani\inst{\ref{inaf_bologna}}
          \and M.~Hamed\inst{\ref{ncbj}}
          \and K.~Ma\l{}ek\inst{\ref{ncbj}, \ref{marseille}}
          \and A.~Durkalec\inst{\ref{ncbj}}
          \and D.~Donevski\inst{\ref{ncbj}, \ref{sissa}, \ref{ifpu}}
          \and S.~Salim\inst{\ref{indiana}}
          \and A.~Iovino\inst{\ref{inaf_brera}}
          \and W.~J.~Pearson\inst{\ref{ncbj}}
          \and M.~Romano\inst{\ref{ncbj}, \ref{oapd}}
          \and M.~Scodeggio\inst{\ref{inaf_milan}}
          }

   \institute{National Centre for Nuclear Research, ul. Pasteura 7, 02-093 Warsaw, Poland\\
              \email{francesco.pistis@ncbj.gov.pl}\label{ncbj}
              \and
              Astronomical Observatory of the Jagiellonian University, Orla 171, 30-001 Cracow, Poland\label{ju}
              \and
              Institute of Astronomy, Faculty of Physics, Astronomy and Informatics, Nicolaus Copernicus University, Grudziądzka 5, 87-100 Toruń, Poland\label{copernicus_torun}
              \and
              INAF - Osservatorio di Astrofisica e Scienza dello Spazio di Bologna, Via Piero Gobetti 93/3, I-40129 Bologna, Italy\label{inaf_bologna}
              \and
              Aix Marseille Univ. CNRS, CNES, LAM, Marseille, France\label{marseille}
              \and
              SISSA, ISAS, Via Bonomea 265, Trieste I-34136, Italy\label{sissa}
              \and
              IFPU, Institute for Fundamental Physics of the Universe, Via Beirut 2, I-34014 Trieste, Italy\label{ifpu}
              \and
              Department of Astronomy, Indiana University, Bloomington, Indiana 47405, USA\label{indiana}
              \and
              INAF - Osservatorio Astronomico di Brera, Via Brera 28, 20122 Milano, via E. Bianchi 46, 23807 Merate, Italy\label{inaf_brera}
              \and
              INAF - Osservatorio Astronomico di Padova, Vicolo dell’Osservatorio 5, I-35122, Padova, Italy\label{oapd}
              \and
              INAF - Istituto di Astrofisica Spaziale e Fisica Cosmica Milano, via Bassini 15, 20133, Milano, Italy\label{inaf_milan}
              }

 \date{Received / Accepted }

 
  \abstract
  {}
    {We investigate the influences on the evolution of the Fundamental Metallicity Relation of different selection criteria.
    }
   {We used $5\,487$ star-forming galaxies at a median redshift $z\approx 0.63$ extracted from the VIMOS Public Extragalactic Redshift Survey (VIPERS) and $143\,774$ comparison galaxies in the local Universe from the GALEX-SDSS-WISE Legacy Catalog.
   We employed two families of methods: parametric and non-parametric. 
    In the parametric approaches, we compared the Fundamental Metallicity Relation projections plagued by observational biases on differently constructed control samples at various redshifts. Then, the metallicity difference between different redshifts in stellar mass-star formation rate bins.
    In the non-parametric approach, we related the metallicity and the normalized specific star formation rate (sSFR). 
    To compare galaxies with the same physical properties, we normalized the median of our samples according to the median sSFR at median redshift $z \approx 0.09$. Then, the galaxies with the same distance from the star-forming main sequence at their respective redshifts are compared when the sSFR is normalized according to the expected values from their respective star-forming main sequence.
   }
   {The methodologies implemented to construct fair, complete samples for studying the mass-metallicity relation and the Fundamental Metallicity Relation produced consistent results showing a small, but still statistically significant evolution of both relations up to $z \approx 0.63$. In particular, we observed a systematic trend where the median metallicity of the sample at $z=0.63$ is lower than that of the local sample at the same stellar mass and star formation rate. The average difference in the metallicity of the low and intermediate redshifts is approximately $1.8$ times the metallicity standard deviation of the median, of the intermediate redshift sample, in stellar mass-star formation rate bins. We confirmed this result using the Kolmogorov-Smirnov test. 
    When we applied the stellar mass-completeness criterion to catalogs, the metallicity difference in redshifts decreased to approximately $0.96$ times the metallicity standard deviation of the median, thus not statistically significant. 
    This result may be dominated by the limited parameter space, being the lower stellar mass galaxies where the difference is larger out from the analysis.
    A careful reading of the results, and their underlying selection criteria, are crucial in studies of the mass-metallicity and fundamental metallicity relations.
   }
   {When studying the mass-metallicity and fundamental metallicity relations, we recommend using the non-parametric approach providing similar results compared to parametric prescriptions, being easier to use and results fair to interpret. The non-parametric methodology provides a convenient way to compare physical properties, with a smaller impact on observational selection biases.
   }


   \authorrunning{F. Pistis et al.}
   \maketitle
%

\section{Introduction}

Stars, embedded in the parent galaxy, produce heavy elements at different stages of their lives, which are released in the gas phase during their evolution and at their death, changing the metallicity, i.e., the amount of heavy elements relative to hydrogen and helium, of the interstellar medium (ISM).
The key processes that impact this abundance during galaxy evolution are inflows, outflows, star formation, and quenching \citep{lilly2013fmr, peng2014evo}, implying that the gas-phase metallicity can be used to trace the star formation history (SFH) and the evolution of the ISM in galaxies.
Measuring the metallicity of galaxies at different epochs constraints and strengthens galaxy formation models \citep{maiolino2019re}, providing information on the early enrichment processes of galaxies and their intergalactic medium (IGM). 

The metallicity is strongly correlated to other galaxy properties, in particular to stellar mass ($M_\star$). The so-called Mass Metallicity Relation \citep[MZR, see, e.g.,][]{tremonti2004origin, savaglio2005gemini, kewley2008metallicity}, has been studied with high precision using, among others, data from the Sloan Digital Sky Survey \citep[SDSS, see, e.g.,][]{tremonti2004origin, mannucci2010fundamental, curti2020mass}, where a positive correlation between metallicity and $M_\star$ was found for star-forming (SF) galaxies. This trend is observed at all redshifts studied so far \citep{lee2006dwarf, maiolino2008amaze, lamareille2009vimos, perez2009mzr, mannucci2010fundamental, zahid2011deep2, cresci2019fundamental, curti2020mass, pistis2022bias}, up to at least $z \sim 3.5$ \citep{maiolino2008amaze}.
Understanding the relation between $M_\star$ and metallicity is crucial when establishing a model of galaxies evolution \citep{lequeux1979chemical}. The origin of the MZR is still debated and different models of galaxy evolution were proposed to explain this relation.
For this aim, these models need to incorporate, among other aspects, the outflow of metal-rich and inflow of metal-poor gas 
\citep{tremonti2004origin, finlator2008origin, dave2010nature, dave2011galaxy, chisholm2018outflow}, the dependence of star formation efficiency on galaxy mass \citep[so-called ``downsizing'' scenario models][]{spitoni2020connection, lilly2013fmr}, and an initial mass function (IMF) which varies with galaxy mass \citep{koppen2007possible, de2018effects, lian2018modelling, lian2018mass}.

The metallicity also shows an anti-correlation with the star formation rate (SFR), as shown by \citet{mannucci2010fundamental, mannucci2011metallicity}, i.e. more strongly star-forming galaxies contain less metals even for galaxies at fixed $M_\star$.
This anti-correlation is in agreement with models including inflows of metal-poor gas, a process that ignites star formation and dilutes the metallicity of the ISM,  and outflows of metal-rich gas, a process that suppresses star formation and removes the metals from the ISM \citep{lilly2013fmr, peng2014evo}.

The combination of this metallicity-SFR relation and MZR is known as the fundamental metallicity relation (FMR).
The FMR has a reduced scatter compared to the MZR and does not show any evolution at least up to $z \sim 2.5$ \citep[][]{mannucci2010fundamental, cresci2019fundamental, curti2020mass}.
The SFR and $M_\star$ are also correlated, which results in a relation known as the main sequence of SF galaxies \citep[MS,][]{brinchmann2004physical, noeske2007ms, elbaz2007ms}.
The MS is then yet another projection of the FMR, where the observed scatter around the MS can be interpreted as a consequence of the inflows of gas resulting in a period of a burst of star formation \citep{abramson2014ms, tacchella2016ms, mitra2017ms} and of the starvation of the galaxy due to the outflows of gas \citep{trussler2020quenching, brownson2022quenching}.
The MS also shows an evolution with redshift \citep{speagle2014ms, schreiber2015ms} with galaxies having higher SFR at a given $M_\star$ at higher redshifts.
Stronger star-forming processes in massive galaxies in the younger Universe may be caused by bigger reservoirs of gas available in a galaxy to form new stars in the earlier cosmic time \citep{tacconi2010sf, scoville2016ism, kokorev2021ism}. 


While most studies about FMR focus on star-forming galaxies, the empirical metallicity calibration for non-SF galaxies based on nebular lines \citep{kumari2019metal} allowed to expand the study of the FMR towards non-SF galaxies \citep{kumari2021fmr}.
The FMR of non-SF galaxies agrees with models that include enhancement of the star formation due to the infall of metal-poor gas and starvation, which prevent the infall stopping the dilution process of metals \citep{kumari2021fmr}.
It remains unclear if the FMR agrees with merger-induced starbursts \citep{gronnow2015merger, bustamante2020mergers} for this class of galaxies.
However, the evolution of the FMR for non-SF galaxies has not yet been studied.

Assuming a regulator system in which the SFR is modulated by the gas reservoir \citep{lilly2013fmr, peng2014evo} allows to model and reproduces the local FMR \citep[][]{mannucci2010fundamental} and MZR \citep[][]{tremonti2004origin, mannucci2010fundamental}.
To model the theoretical evolution of the FMR, this model considers i) the evolution in cosmic time of the specific SFR (sSFR, defined as the ratio between SFR and $M_\star$) corresponding to the halo growth, ii) the gas-phase metallicities over the galaxy population, and iii) the stellar to dark matter mass ratio of the halos.

The comparison of galaxy samples at different redshifts involves many difficulties, including: i) different selection effects \citep[e.g., survey limit, S/N of spectra][]{pistis2022bias}; ii) different rest-frame observed; iii) different selection criteria for SF galaxies and for estimation of metallicity; iv) different metallicity calibrators based on different sets of emission lines or different methodology of calibration \citep{kewley2008metallicity}; v) small samples available at higher $z$ (as seen from Table~\ref{tab:samplesizes}); vi) different methods of selecting galaxies for comparison; vii) different methods of comparison (to be addressed in this study).
\begin{table}
\caption{Samples used in different studies of the FMR.}              
\label{tab:samplesizes}      
\centering                                      
\begin{tabular*}{\columnwidth}{l @{\extracolsep{\fill}} c c c}          
\hline    
\hline
\noalign{\smallskip}
Survey & Reference & $z$ & Sample size \\    
\hline                                   
\noalign{\smallskip}
    SDDS & 1 & $0.07$ -- $0.30$ & $141\,825$  \\
    Collection  & 1 & $0.5$ -- $2.5$ & $182$  \\
    LSD, AMAZE  & 1 & $3$ -- $4$ & $16$  \\
    zCOSMOS  & 2 & $0.2$ -- $0.8$ & $334$  \\
    SDSS  & 3 & $0.005$ -- $0.25$ & $177\,071$  \\
    MOSDEF   & 4 & $\sim 2.3$ & $87$  \\
    zCOSMOS   & 5 & $05$ -- $0.9$ & $39$  \\
    KBSS  & 6 & $\sim 2.3$ & $130$  \\
    eBOSS  & 7 & $0.6$ -- $0.9$ & $35$  \\
    CALIFA  & 8 & $0.005$ -- $0.03$ & $612$  \\
    SDSS-IV MANGA  & 8 & $0.03$ -- $0.17$ & $2730$  \\
    SDSS  & 9 & $0.027$ -- $0.3$ & $153\,452$  \\
    SDSS  & 10 & $0.005$ -- $0.2$ & $68\,942$  \\
    MOSDEF & 11 & $2.3$ -- $3.3$ & $\sim 450$ \\
    KBSS & 12 & $2$ -- $3$ & $150$ \\
    SDSS  & 13 & $0.027$ -- $0.27$ & $156\,018$  \\
    VIPERS & 13 & $0.5$ -- $0.8$ & \textbf{4\,772}  \\
    SDSS  & This work & $0.027$ -- $0.27$ & $143\,774$  \\
    VIPERS  & This work & $0.5$ -- $0.8$ & \textbf{5\,487}  \\
    \noalign{\smallskip}
    \hline
\end{tabular*}
\tablebib{
(1)~\citet{mannucci2010fundamental}; (2)~\citet{cresci2012zcosmos}; (3)~\citet{yates2012relation}; (4)~\citet{sanders2015mosdef}; (5)~\citet{maier2015fmr}; (6)~\citet{salim2015mass}; (7)~\citet{gao2018mass}; (8)~\citet{cresci2019fundamental}; (9)~\citet{curti2020mass}; (10)~\citet{bustamante2020mergers}; (11)~\citet{sanders2021mosdefevo}; (12)~\citet{strom2022kbss}; (13)~\citet{pistis2022bias}.}
\end{table}

In order to investigate if the FMR is really fundamental, or depends on (redshift variant) methods of comparing the samples, we make use of the unprecedented statistics of the VIMOS Public Extragalactic Redshift Survey (VIPERS) to study and compare, for the first time with a high statistical significance, the MZR and FMR at median $z \sim 0.63$ and at median $z \sim 0.09$.
In this paper, we use a variety of methods of comparison in order to determine how strong the conclusions are depending on the method used.

We apply the following methods: i) a family of parametric methods, based on the direct comparison of different projections of the FMR, ii) a non-parametric method based on \citet{salim2014critical, salim2015mass}.
The non-parametric method is based on the comparison between the metallicity and the normalized sSFR in different $M_\star$ bins.
The choice of normalization for the sSFR in the non-parametric method allows us to choose the properties that will be compared between the samples.


The paper is organized as follows.
In Sect.~\ref{sec:data} we describe the main samples with their initial data selection.
In Sect.~\ref{sec:samp_building} we describe the construction of the so-called control samples, i.e., sub-samples with specific properties cross-matched at different redshift ranges.
In sect.~\ref{sec:methods} we present two families of methods to compare low and intermediate redshift data.
In Sect.~\ref{sec:comparison} we present the comparison of the samples.
In Sect.~\ref{sect:mzrevo}, we analyze the evolution of the MZR and metallicity-SFR relation.
In Sect.~\ref{sec:disc} and \ref{sec:concl}, we discuss the results and present the final conclusions.
The cosmological parameters adopted in this paper are: $H_0 = 70\ \text{km} \, \text{s}^{-1} \, \text{Mpc}^{-1}$; $\Omega_M = 0.3$; $\Omega_\Lambda = 0.7$; we assume a \cite{chabrier2003galactic} IMF.

\section{Main data samples}\label{sec:data}

We use spectroscopic data from two surveys: VIPERS (${0.5 < z < 1.2}$) and SDSS ($ 0 < z < 0.3$).
We also make use of the VIMOS VLT Deep Survey \citep[VVDS,][]{lefevre2013vvds} to validate the spectroscopic measurements performed on the VIPERS sample and to validate the shape of the MZR obtained with VIPERS.
The main selection steps are described below and a more detailed description of this selection can be found in \cite{pistis2022bias}.
Table~\ref{tab:sel} summarizes the data selection at each step.
\begin{table}
\caption{Steps of data selection and size of VIPERS and SDSS samples after each selection step.}              
\label{tab:sel}      
\centering                                      
\begin{tabular*}{\columnwidth}{l @{\extracolsep{\fill}} c c}          
\hline    
\hline
\noalign{\smallskip}
Selection & VIPERS & SDSS \\    
\hline                                   
\noalign{\smallskip}
    Spectroscopic sample & $88\,340$ & $927\,552$\\      
    Sample with properties & $39\,204$ & $601\,082$\\
    $\text{H}\alpha$ $\left( \mathrm{S/N} > 15 \right)$ & -- & $299\,070$ \\
    $\text{H}\beta$ $\left( \mathrm{S/N} > 3 \right)$ & $20\,545$ & $296\,027$ \\
    Redshift & -- & $279\,851$\\
    Redshift flag & $20\,545$ & --\\
    $\mathrm{E} \left( \mathrm{B} - \mathrm{V} \right) < 0.8$ & $20\,545$ & $267\,922$\\
    Line flags & $9\,290$ & --\\
    BPT diagram & $6\,251$ & $158\,416$\\
    Upper branch & $6\,116$ & $157\,404$\\
    Metallicity calibration ($\mathrm{Z} > 8.4$) & $6\,018$ & $157\,223$\\
    Error on metallicity ($\sigma_\mathrm{Z} < 0.3$) & $5\,776$ & $151\,345$\\
    sSFR ($>$5th percentile) & $5\,487$ & $143\,774$\\
    \noalign{\smallskip}
    \hline
\end{tabular*}
\tablefoot{
Selection on S/N on   $\text{H}\alpha$ was applied only for the SDSS as this line is not visible in the VIPERS spectra. The redshift range of the VIPERS sample ($0.48<z<0.8$) is a natural consequence of the requirement of having the lines $\text{H}\beta$, $\left[ \ion{O}{ii} \right]\lambda 3727$, $\left[ \ion{O}{iii}\ \right]\lambda 4959 $, and $\left[ \ion{O}{iii} \right]\lambda 5007$ in the spectral range. In the SDSS, we do not use the redshift nor line flags, but the spectral quality is assured by the high S/N of $\text{H}\alpha$ and we limit the redshift at $z \geq 0.027$ to include $\left[ \ion{O}{ii} \right]$ doublet in the spectra.
}
\end{table}

\subsection{VIPERS sample}

We build a sample of galaxies at intermediate redshift by cross-matching two catalogs.
The first one is the Public Data Release-2 (PDR2) spectroscopic catalog of VIPERS which contains $88\,340$ galaxies \citep{scodeggio2018vimos}.
The target selection requires brighter sources than $i_\mathrm{AB} = 22.5$ and a pre-selection in the $(u - g)$ and $(r - i)$ color-color diagram is made to remove galaxies at low redshift.
The VIPERS has a spectral resolution of $R \sim 250$, which makes it possible to examine specific spectroscopic characteristics of galaxies having a wavelength coverage ranging from 5500 to 9500 \AA.
The data reduction pipeline and redshift quality system are described in \cite{garilli2014vimos}. 
This catalog contains the redshifts and flags used to define the confidence level down to $99$\% ($3.0 \leq z_\text{flag} \leq 4.5$).

This catalog is then supplemented with the new spectroscopic measurements of fluxes and equivalent widths (EWs) of the emission lines of interest in this paper: $\text{H}\beta$, $\left[ \ion{O}{ii} \right]\lambda 3727$, $\left[ \ion{O}{iii} \right]\lambda 4959$, and $\left[ \ion{O}{iii} \right]\lambda 5007$.
The VIPERS spectra are analyzed with the penalized pixel fitting code \citep[pPXF,][]{cappellari2004fit, cappellari2017improving, cappellari2022full}, which allows fitting both the stellar and gas components via full spectrum fitting.
After shifting the observed spectra to the rest frame and masking out the emission lines, the stellar component of the spectra is fitted with a linear combination of stellar templates from the MILES library \citep{vazdekis2010miles} after being convolved to the same spectral resolution as the observations.
The gas component is then fitted with a single Gaussian for each emission line, giving the integrated fluxes and their errors as a result.
In order to have a better estimation of the error, the error given by pPXF is then multiplied by the $\chi^2_\text{red}$ of the fit under the emission line.

In order to estimate the EWs from the pPXF results, it is useful to normalize to one the spectral continuum of each galaxy.
We build the normalized spectrum by dividing the best fit by the stellar component of the fit.
The spectra are then analyzed with specutils, an astropy package for spectroscopy \citep{astropy2013, astropy2018, price2018astropy} in a range of $\pm 1.06$ full-width half maximum (FWHM), which is equivalent to $5$ standard deviation of the Gaussian fit \citep{vietri2022agn}, around the centroid of the emission line to estimate the EW and its uncertainty.

For this new catalog of spectroscopic measurements, we adopt the same flag system used in the VIPERS catalogs \citep{garilli2010ez, figueira2022sfr, pistis2022bias} in the form of a 4 digits number $xyzt$.
The $x$-value is equal to $1$ if the difference between the centroid of the fit and the centroid of the observed data is less or equal to $7$~{\AA} (equivalent to $1$ pixel on the VIMOS spectrograph), else its value is $0$.
The $y$-value is equal to $1$ if the FWHM is in the range $7$--$22$~{\AA} equivalent to $1$--$3$ pixels of the spectrograph, else its value is $0$.
The $z$-value is equal to $1$ if the difference between the peak of the data and the fit is less than $30\%$, else its value is $0$.
Finally, the $t$-value is equal to $2$ if the signal-to-noise ratio (S/N) for the EW is greater than $3.5$ or the S/N for the flux is at least $8$, its value is $1$ if the S/N for the EW is at least $3$ or the S/N for the flux is at least $7$, else its value is $0$.

As a check for the quality of the measurements, we compare our results with a catalog from the VVDS \citep{lamareille2009vimos}, as both VIPERS and VVDS used the same instrumental configuration of VIMOS spectrograph \citep{lefevre2003vvdsdesign}.
We select all galaxies from the VVDS sample covering the same $z$ and $z_\mathrm{flag}$ ranges as the VIPERS sample. We also limited the VVDS sample to the same magnitude as VIPERS, $i_\text{AB} < 22.5$, removing deeper observations included in VVDS. The resulting catalog has similar characteristics as VIPERS, both from an instrumental and continuum treatment point of view.
For the comparison, we select galaxies from the VIPERS sample with a minimum flag of $1110$ for the lines $\text{H}\beta$, $\left[ \ion{O}{ii} \right]\lambda 3727$, $\left[ \ion{O}{iii} \right]\lambda 4959$, and $\left[ \ion{O}{iii} \right]\lambda 5007$.

The VVDS sample has been analyzed with the {platefit\textunderscore vimos} pipeline \citep{lamareille2009vimos}, which is the adapted version of platefit used to analyze high-resolution SDSS spectra \citep{tremonti2004origin, brinchmann2004physical}.
To make the comparison between VIPERS and VVDS consistent, the same treatment of the continuum and stellar part adopted by {platefit\textunderscore vimos} was used in pPXF, fitting separately gas and stellar spectra.

\begin{figure}
    \centering
    \resizebox{\hsize}{!}{\includegraphics{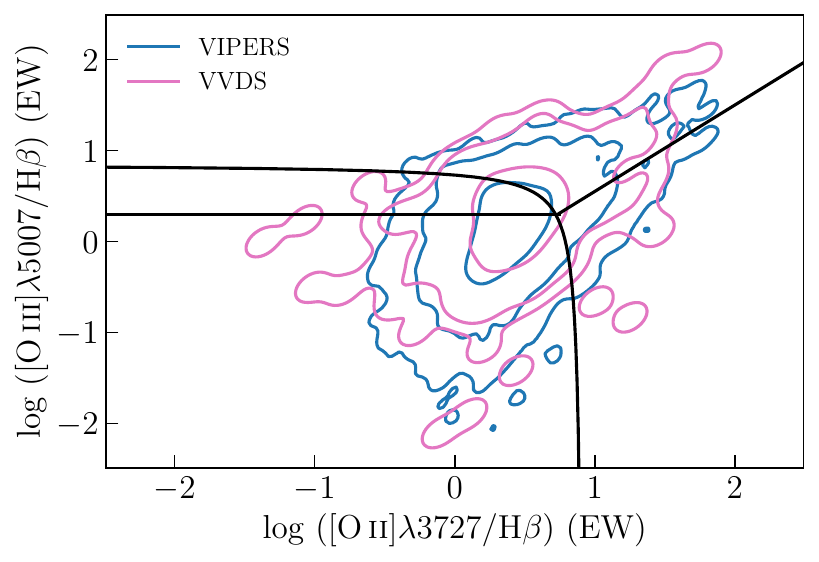}}
    \caption{``Blue'' BPT diagram \citep{lamareille2010spectral} for VIPERS (blue) and VVDS (pink) samples. Contours present $1$, $2$, and $3$ standard deviation levels of the distributions. The black lines are the thresholds defined in \cite{lamareille2010spectral} to divide different galaxy types (SF, AGN, and LINERs).}
    \label{fig:bpt_comp}
\end{figure}

Figure~\ref{fig:bpt_comp} shows the $1$, $2$, and $3$ standard deviation levels of the surface density distribution contours in the BPT diagram \citep{BPT81} for VIPERS and VVDS samples.
Both samples show very compatible distributions.
We report further analysis to validate the new line measurements in Appendix~\ref{app:val} and the S/N distributions of the main emission lines in Appendix~\ref{app:hists}.

The catalog of line measurements obtained by this procedure is then cross-matched with the physical properties (in particular $M_\star$) catalog for VIPERS galaxies \citep{turner2021unsupervised} measured via the spectral energy distribution (SED) fitting with the Code Investigating GALaxy Emission \citep[CIGALE,][]{burgarella2005star, noll2009analysis, boquien2019cigale}.
The construction of the catalog reduces the sample to $39\,204$ galaxies since the catalog with the physical properties is not complete in the W4 field.

We select galaxies with a $3.0 \le z_\mathrm{flag} \le 4.5$, $\mathrm{S/N} > 3$ for $\mathrm{H}\beta$ flux , $\mathrm{E} \left( \mathrm{B} - \mathrm{V} \right) < 0.8$, \citep{curti2020mass}, and minimum line flags of $1110$ for the lines $\text{H}\beta$, $\left[ \ion{O}{ii} \right]\lambda 3727$, $\left[ \ion{O}{iii} \right]\lambda 4959$, and $\left[ \ion{O}{iii} \right]\lambda 5007$.
The S/N selection is not applied to oxygen lines since it was shown to bias the FMR and its projections \citep{yates2012relation, kashino2016fmr, telford2016fmrsystematic, cresci2019fundamental, pistis2022bias}.
Having the emission lines ($\mathrm{H}\beta$, $\left[ \ion{O}{ii} \right ]\lambda 3727$, $\left[ \ion{O}{iii} \right ]\lambda 4959$, and $\left[ \ion{O}{iii} \right ]\lambda 5007$) limits the redshift range to $0.48 < z < 0.80$.
These selections limit the sample size to $9\,290$ galaxies.

To select SF galaxies and exclude LINERS and Seyfert galaxies from the sample, we use the lines ratio classification of the BPT diagram.
The original line ratio diagram to select different types of active galaxies uses $\left[ \ion{O}{iii} \right]\lambda 5007$, $\left[ \ion{N}{ii} \right ]\lambda 6584$, $\left[ \ion{S}{ii} \right ]\lambda 6717, 6731$, $\mathrm{H}\alpha$, and $\mathrm{H}\beta$.
Because of the redshift range covered by VIPERS ($z>0.5$ for most of its sample) and the VIMOS spectrograph wavelength coverage ($5500$ -- $9500$ {\AA}), the emission lines necessary for the original BPT diagram \citep{kauffmann2003stellar} are not detected ($\left[ \ion{N}{ii} \right ]\lambda 6584$ and $\left[ \ion{S}{ii} \right ]\lambda 6717, 6731$) or detected only for few galaxies ($\mathrm{H}\alpha$). For this sample, it is necessary to use the so-called ``blue'' BPT diagram  \citep{lamareille2010spectral}.
This version of the BPT diagram can be used for a higher redshift range and uses the $\left[ \ion{O}{ii} \right]\lambda 3727$, $\left[ \ion{O}{iii} \right]\lambda\lambda 4959, 5007$, and $\text{H}\beta$ spectral lines.
The selection of SF galaxies based on the \cite{lamareille2010spectral} recipe reduces the sample to $6\,251$ galaxies.

The computations of the properties starting from the line fluxes are done via bootstrap.
For each galaxy, we generate the fluxes and attenuation as random numbers with Gaussian distribution centered at the ``observed'' values and with standard deviations equal to estimated errors.
This process is repeated $1\,000$ times for each galaxy.
We then calculate all the derived quantities ($R_{23}$, SFRs, and metallicity) as the median and the error as the standard deviation of the distributions generated during the bootstrap.

To correct the flux of emission lines for attenuation, we used the attenuation in the V-band ($A_\text{V}$, see Appendix~\ref{app:av} for the estimation of this property) provided by the fit of the SED via the CIGALE code, corrected by the f-factor \citep[$\mathrm{f} = 0.57$,][]{rodriguez2022attenuation} to pass from stellar to nebular $A_\text{V}$, and a \cite{cardelli1989relationship} attenuation law with $R_V=3.1$.

The SFR is computed from the $\left[ \ion{O}{ii} \right]$ luminosity \citep{kennicutt1998star} transformed into a \cite{chabrier2003galactic} IMF:
\begin{equation}
    \mathrm{SFR}_{\left[ \ion{O}{ii} \right]} = 8.2 \times 10 ^{-42}\, \mathrm{L}_{\left[ \ion{O}{ii} \right]}
\end{equation}
where $\mathrm{L}_{\left[ \ion{O}{ii} \right]}$ is expressed in unit of $\mathrm{erg}\, \mathrm{s}^{-1}$.
The $\left[ \ion{O}{ii} \right]$ luminosity is previously corrected for attenuation.


The metallicity is estimated via the calibration of \cite{tremonti2004origin} based on the $R_{23}$ parameter \citep{pagel1979hii}:
\begin{equation}
        R_{23} = {\frac{\left[ \ion{O}{ii} \right]\lambda 3727 + \left[ \ion{O}{iii} \right]\lambda\lambda 4959, 5007}{\text{H}\beta}},
\end{equation}
and
\begin{equation}
    12 + \log \left( \mathrm{O/H} \right) = 9.185 - 0.313 x - 0.264 x^2 - 0.321 x^3,
\end{equation}
where $x \equiv \log R_{23}$.
This calibration is valid only for the upper branch of the double-valued $R_{23}$ abundance relation.
We select the upper branch according to the threshold ${\left[ \ion{O}{iii} \right] \lambda 5007 / \left[ \ion{O}{ii} \right] \lambda 3727 < 2}$ \citep{nagao2006gas}.
This calibration is valid only for ${12 + \log \left( \text{O/H} \right) > 8.4}$. Consequently, we remove galaxies for which we obtained a lower value of $12 + \log \left( \text{O/H} \right)$.
We also remove all galaxies with uncertainties on metallicity (estimated by bootstrap method) larger than $0.3$ dex.
After these steps, our final intermediate redshift sample is reduced to $6\,018$ SF galaxies with a median redshift of $z \sim 0.63$.

\subsection{SDSS sample}\label{sec:sdss}

We constructed the sample at low redshift by cross-matching two catalogs: the MPA/JHU catalog\footnote{\url{https://wwwmpa.mpa-garching.mpg.de/SDSS/DR7/}} based on SDSS DR7,  composed of $927\,552$ galaxies with spectroscopic redshift (in the range $0.0< z < 0.7$) and line fluxes \citep{kauffmann2003stellar, brinchmann2004physical, tremonti2004origin} and the A2.1 version of the GALEX-SDSS-WISE Legacy Catalog\footnote{\url{https://salims.pages.iu.edu/gswlc/}} \citep[GSWLC-2,][]{salim2016galex, salim2018dust} with $640\,659$ galaxies at $z < 0.3$ based on SDSS DR10 \citep{anh2014sdss}, and containing physical properties ($M_\star$, SFR, and absolute magnitudes) obtained through SED fitting with CIGALE.
The cross-matched sample contains $601\,082$ galaxies.

To obtain the sample we will use in our analysis, we followed the data selection introduced by \cite{curti2020mass}. We apply an S/N limit of 15 and 3 for $\text{H}\alpha$ and $\text{H}\beta$ spectral lines, respectively, and corrected all emission lines for attenuation using the Balmer decrement, assuming the case B recombination \citep[$\text{H}\alpha / \text{H}\beta = 2.87$,][]{baker1938balmer} and adopting the \cite{cardelli1989relationship} law with $R_V = 3.1$, correcting again the $\mathrm{A}_\mathrm{V}$ by the f-factor.
We limited the SDSS sample to $z\ge 0.027$ to include the $\left[ \ion{O}{ii} \right]$ line in the observed spectra and removed all galaxies showing high extinction, i.e. with values of $\text{E} \left( \text{B} - \text{V} \right)$ higher than $0.8$.
After these selections, the cross-matched sample is composed of $267\,922$ galaxies.

To have a homogeneous selection for both the SDSS and VIPERS samples, we selected SF galaxies using the same ``blue'' BPT diagram \citep{lamareille2010spectral} as for the case of galaxies at intermediate redshift.
As shown in \cite{pistis2022bias}, the use of this diagram does not significantly bias the analysis of the FMR.
The SF selection further reduces the sample to $158\,416$ sources.

The physical properties are computed with the same bootstrap method and calibrations used for the VIPERS sample.
Because the fiber system of the SDSS survey measures the light with different spatial coverage, the SFR needs a correction to take into account the fiber aperture \citep{hopkins2003star}:
\begin{equation}
\mathrm{SFR}_\mathrm{final} = \mathrm{SFR}_{\left[ \ion{O}{ii} \right]} \times 10^{-0.4 \left( u_\mathrm{Petro} - u_\mathrm{fiber} \right)},
\end{equation}
where $u_\mathrm{Petro}$ and $u_\mathrm{fiber}$ are the modified forms of the Petrosian magnitude \citep{petrosian1976surface} and the magnitude measured within the aperture of the spectroscopic fiber, respectively.
We also removed all galaxies with $12 + \log \left( \text{O/H} \right) < 8.4$ and a metallicity error larger than $0.3$ dex.

From this sample, we also removed all sources with $\log M_\star  \left[M_\sun\right] < 7$, $\log \text{SFR} \left[M_\sun/\text{yr}\right] <-10$ to exclude low-mass galaxies, not present in the VIPERS sample, as well as all remaining passive objects.
Finally, we removed galaxies that did not possess a reliable rest-frame blue magnitude by applying a cut at $M_\text{B} >-24$\footnote{A $\text{B}$ magnitude flagged as $-99$ in the GSWLC-2 catalog indicates a failure in the estimation of the parameter during the fit procedure.}.
All those selections removed galaxies with bad SED fit results and reduced the total number of local SF galaxies to $155\,893$ with a median redshift $z \sim 0.09$.

\subsection{Homogeneous star-forming main sequence}

To build one of the control samples used in the following part of this work (called a distance control sample, see Sect.~\ref{sec:samp_building} and Sect.~\ref{sec:nonparam}) we need properties that are derived from the star-forming MS.
For this purpose, we need a homogeneous definition of the MS at low and intermediate redshift.

To take into account the MS evolution in the wide redshift range of the VIPERS sample ($0.5 \le z \le 0.8$), we divide this sample into three redshift bins: ${0.5 \le z < 0.6}$, ${0.6 \leq z \leq 0.7}$, and ${0.7 < z \le 0.8}$.
We perform a linear fit of the SFR-$M_{\star}$ distribution separately in all three redshift bins.
In the next step, we perform another linear fit to the redshift dependence of the MS parameters (slopes and intercepts) based on these three fits.
In this way, we obtain the following redshift-dependent MS relation:
\begin{equation}
    \log \text{SFR}_\text{MS}^\text{VIPERS} \left( z , M_\star \right) = \alpha \left( z \right) \log M_\star  + \beta \left( z \right),
\end{equation}
with ${\alpha \left( z \right) = -0.04\ z + 0.70}$ and ${\beta \left( z \right) = 1.37\ z - 6.44}$, valid in the redshift range $0.5<z<0.8$. A detailed analysis of the VIPERS MS has been published by \cite{pearson2023msvipers}.

The redshift range explored by SDSS is small enough so that a single linear fit of the SFR-M$_{\star}$ relation is sufficient for our analysis.
The SDSS MS relation is given by:
\begin{equation}
    \log \text{SFR}_\text{MS}^\text{SDSS} \left( M_\star \right) = 0.92 \log M_\star - 8.50.
\end{equation}

\begin{figure}
    \centering
    \resizebox{\hsize}{!}{\includegraphics{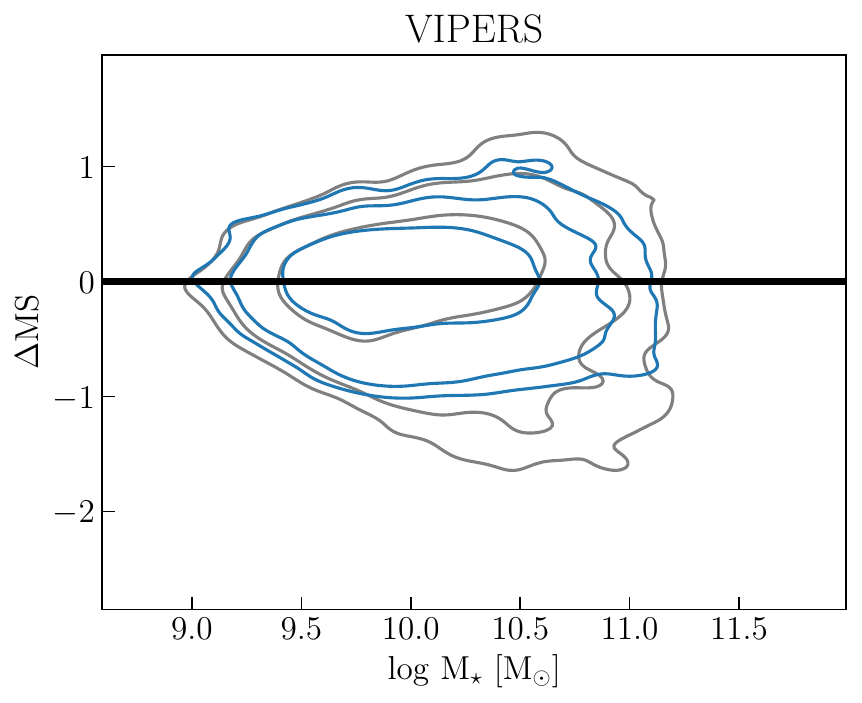}\includegraphics{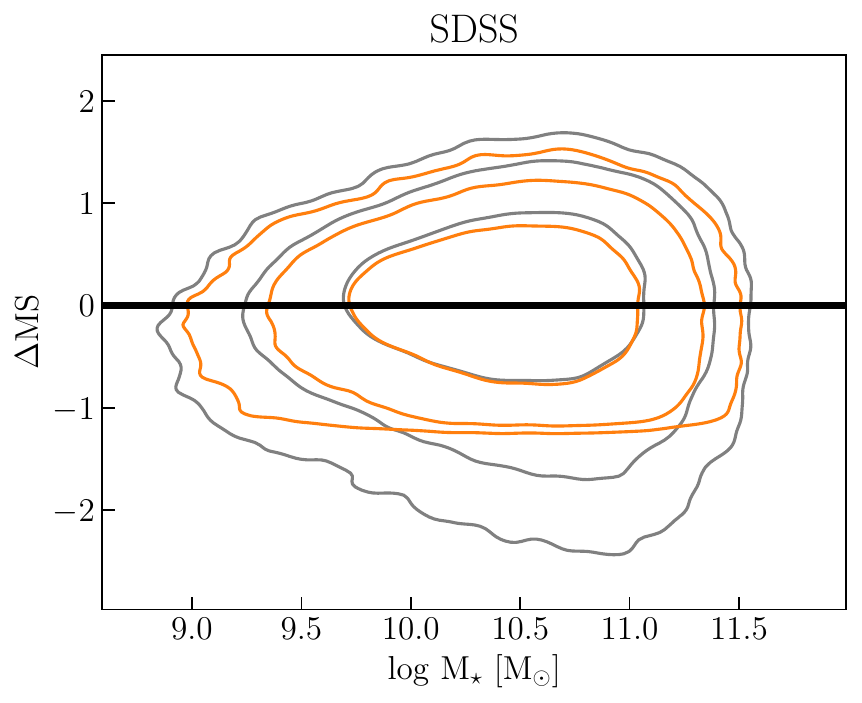}}
    \resizebox{\hsize}{!}{\includegraphics{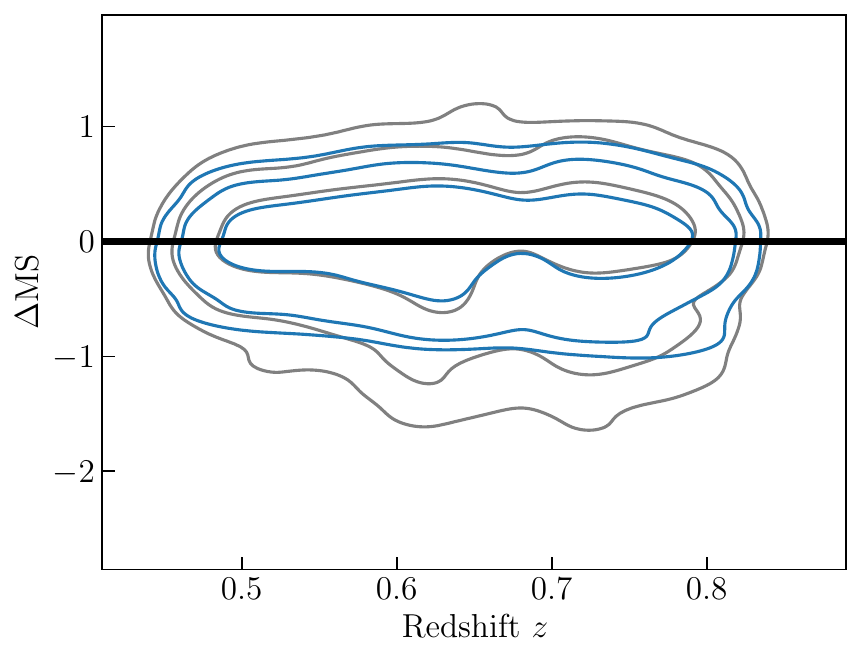}\includegraphics{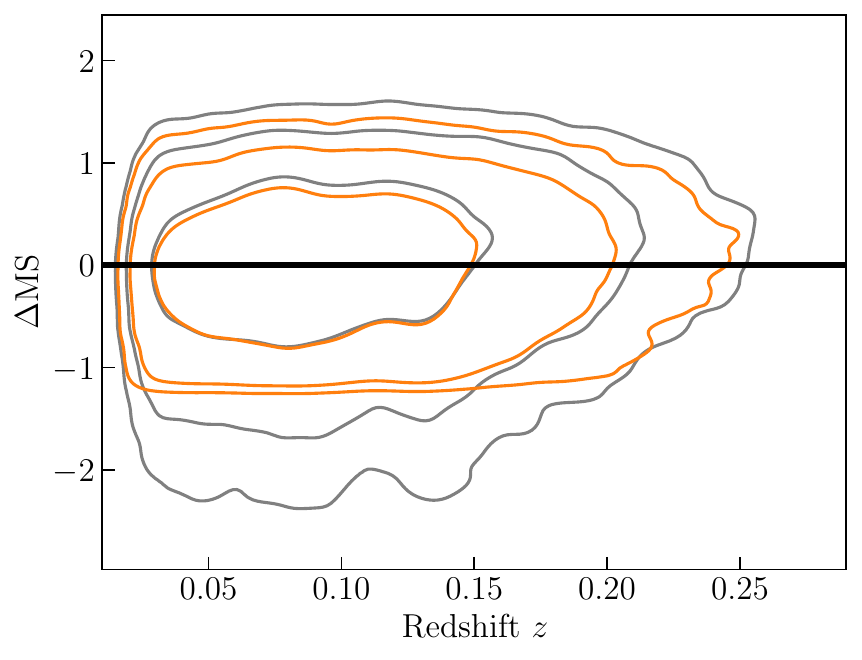}}
    \caption{Scatter around the MS as a function of $M_\star$ (upper panels) and redshift (bottom panels) for VIPERS (blue) and SDSS (orange) samples. The grey contours show the scatter around the MS for the samples before the cut in sSFR. Contours show the $1\sigma$, $2\sigma$, and $3\sigma$ levels of the distributions.}
    \label{fig:delta_ms_vipers}
\end{figure}

\begin{figure}
    \centering
    \resizebox{\hsize}{!}{\includegraphics{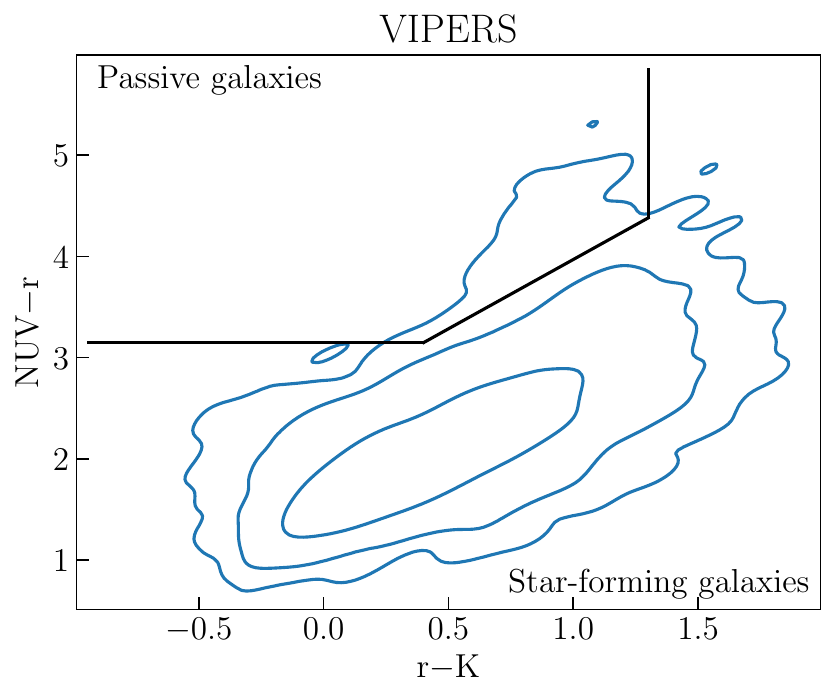}\includegraphics{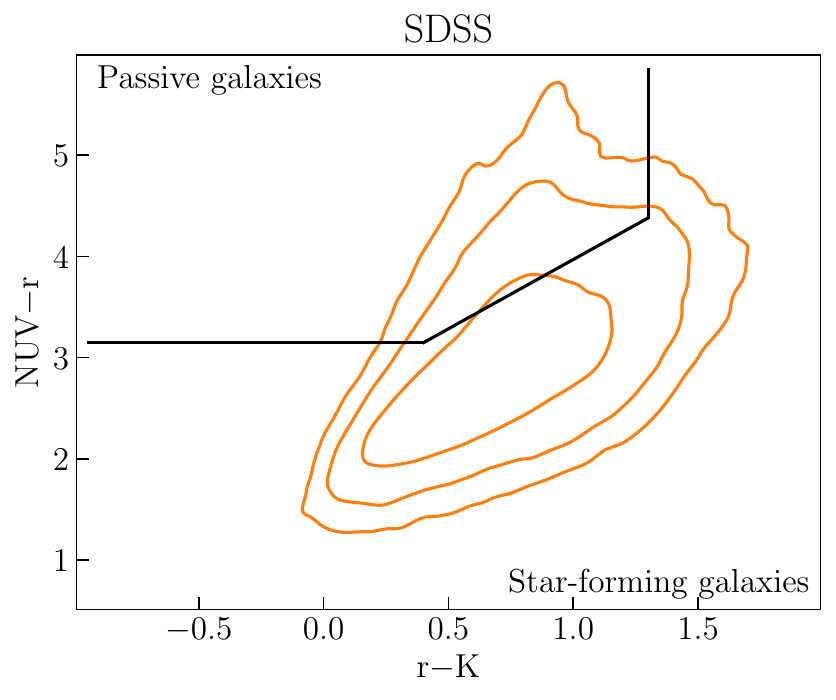}}
    \caption{Distribution in the NUVrK diagram for VIPERS (blue) and SDSS (orange) samples. Contours show the $1\sigma$, $2\sigma$, and $3\sigma$ levels of the distributions.}
    \label{fig:NUVrK}
\end{figure}

Figure~\ref{fig:delta_ms_vipers} shows the scatter around the MS as a function of $M_\star$ and redshift for VIPERS and SDSS samples, demonstrating a lack of any statistically significant trend with either $M_\star$ and redshift for both samples.
Both samples show a similar standard deviation level of $1\sigma$ of the distributions.
The low redshift (SDSS) sample has a bigger scatter around the MS compared to the intermediate redshift (VIPERS) sample, especially at high-$M_\star$ and low redshift.
It can be caused by contamination with galaxies moving between active and passive stages that are not removed by the blue BPT diagram used to select SF galaxies.
Checking the NUVrK diagram \citep[see Fig.~\ref{fig:NUVrK},][]{davidzon2016env}, we find two remaining passive galaxies in the VIPERS sample and $1\,780$ ($\sim 0.1 \%$ of the sample) remaining passive galaxies in the SDSS sample.

Because of the presence of passive galaxies especially at large $M_\star$ that are not removed by the data selection, we add an additional cut in sSFR removing all galaxies below the 5th percentile in the sSFR distribution, reducing the VIPERS sample to $5\,717$ SF galaxies and the SDSS sample to $148\,098$ SF galaxies.
Then, we perform again the fit of the MS finding:
\begin{equation}
    \begin{split}
        \alpha \left( z \right) &= 0.51\  z + 0.56 \\
        \beta \left( z \right) &= -4.24\  z - 4.95
    \end{split}
\end{equation}
for VIPERS sample, and
\begin{equation}
    \log \text{SFR}_\text{MS}^\text{SDSS} \left( M_\star \right) = 0.98 \log M_\star - 9.11
\end{equation}
for the SDSS sample.
These new MS fits result to be consistent with those found by \cite{whitaker2012ms, lee2015msturnover}.

\section{Control samples}\label{sec:samp_building}

In order to study FMR evolution and the robustness of the result against different observational biases, two control samples have been built from our main sample of SDSS galaxies (see  Sect.~\ref{sec:sdss}). These are:
\begin{itemize} 
\item the property-control sample (hereafter p-control) having similar $M_\star$ and SFR as VIPERS galaxies;
\item the distance-control sample (hereafter d-control) having the same relative distance from the MS as VIPERS.
\end{itemize}
If the FMR is fundamental --- the metallicity depends only on the given $M_\star$ and SFR and not from the specific stage of evolution for each galaxy --- and does not evolve with redshift, the differences in metallicity in the FMR and its projections between the VIPERS sample and the four SDSS samples (all selected SF galaxies, p-control and d-control samples) should be independent of each other.
If the metallicity differences in the FMR and its projections are independent of the methodology of cross-matching, the differences depend only on the physical properties ($M_\star$ and SFR) of galaxies.


To build the p-control sample, we select for each VIPERS galaxy the three closest SDSS galaxies on the $M_\star$-SFR plane, up to $0.1$~dex.
In this way, we avoid unbalancing the galaxy distribution towards the region of the MS where SDSS is much denser compared to the VIPERS sample.
This selection reduces the p-control SDSS sample to $12\,053$ galaxies ($\sim 8\%$ of the sample) after removing galaxies counted multiple times.

The MS relations for both SDSS and VIPERS were defined in the same homogeneous way, so it can be used to build the d-control sample.
To create this sample, for each VIPERS galaxy, we move vertically over the MS defined by the SDSS sample.
We then simulate the scatter around the local MS for the d-control sample adding a random number with Gaussian distribution centered at ${\mu = 0}$ and width $\sigma$ (standard deviation of SFR in a $0.1$ dex width bin around $M_\star$ of VIPERS sample):
\begin{equation}
    \log \text{SFR}_\text{d-control} \left( M_\star \right)= \log \text{SFR}_\text{MS}^\text{SDSS} \left(M_\star \right) + N \left( \mu, \sigma \right).
\end{equation}
We then proceed to select a maximum of three closest SDSS galaxies in a radius of $0.1$ dex in the $M_\star$-SFR plane.
The d-control sample constructed in this way is composed of $14\,475$ SDSS galaxies  ($\sim 10\%$ of the sample) with the same distance from MS as galaxies in the VIPERS sample.

To summarize, we have four samples: i) the VIPERS sample at intermediate redshift and ii) the SDSS sample at low redshift, both built using similar selection criteria and consisting of SF galaxies and two control samples obtained starting from the SDSS sample and selecting subsamples that simulate at lower redshift the properties of the VIPERS galaxy sample: iii) the p-control (mimic the physical properties $M_\star$ and SFR), iv) d-control samples (mimic the scatter around the MS).
In the parametric method, we use both main and control samples.
In the non-parametric method, instead, we use only the main samples.

Starting from the VIPERS sample, we build a sub-sample by applying the condition to have a stellar mass-complete sample. Following \citet{davidzon2016env}, we adopt the mass threshold as ${\log M_\star \left[ M_\sun \right] = 10.18}$ for ${0.51 < z  \le 0.65}$,  ${\log M_\star \left[ M_\sun \right] = 10.47}$ for ${0.65 < z \le 0.8}$, and  ${\log M_\star \left[ M_\sun \right] = 10.66}$ for ${0.8 < z  \le 0.9}$.
At the same time, we consider the sub-sample composed of the galaxies not included in the stellar mass-complete sample, calling this sample VIPERS not mass-complete.
These two sub-samples will be used in the Sect.~\ref{sect:mzrevo}.

\subsection{General properties of VIPERS and SDSS main samples and sub-samples}\label{sec:dists}

\begin{figure*}
    \centering
    \resizebox{\hsize}{!}{\includegraphics{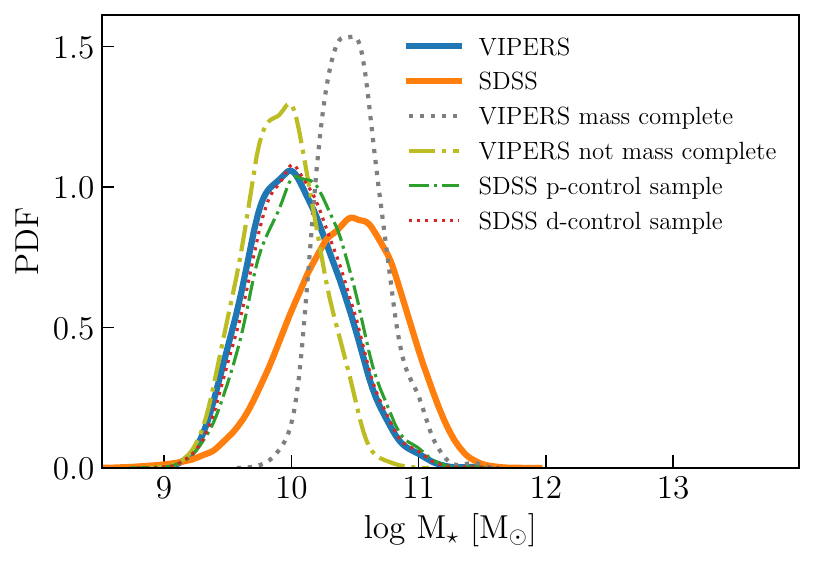}\includegraphics{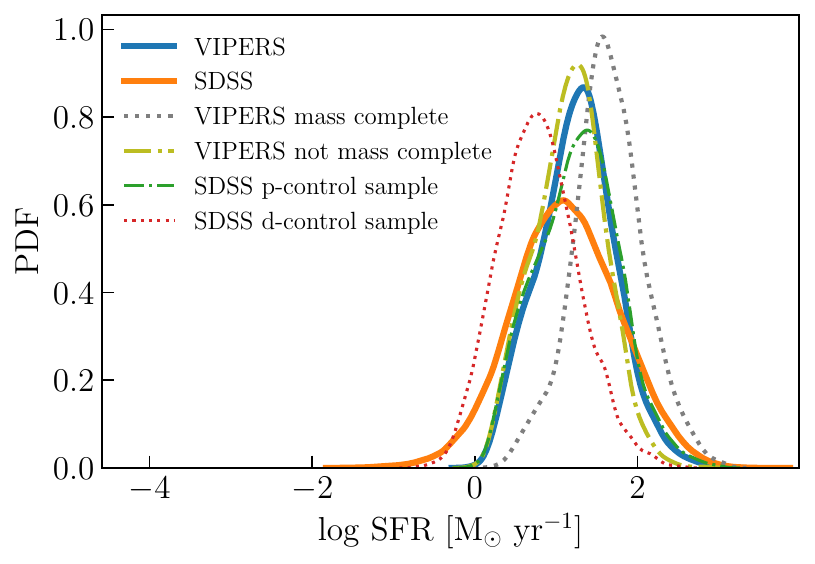}\includegraphics{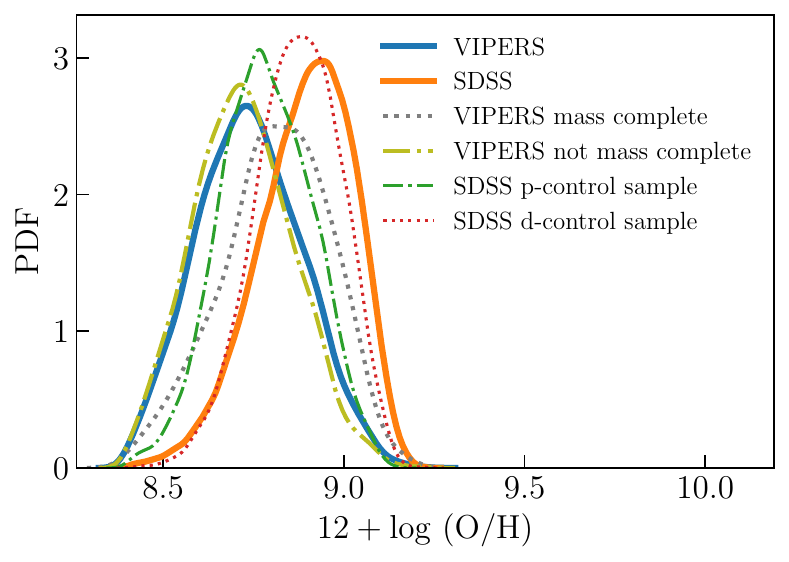}}
    \resizebox{\hsize}{!}{\includegraphics{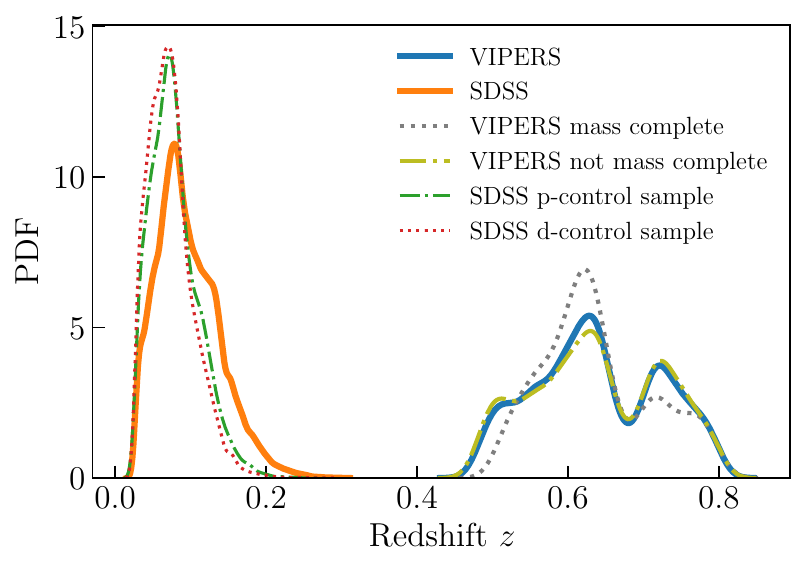}\includegraphics{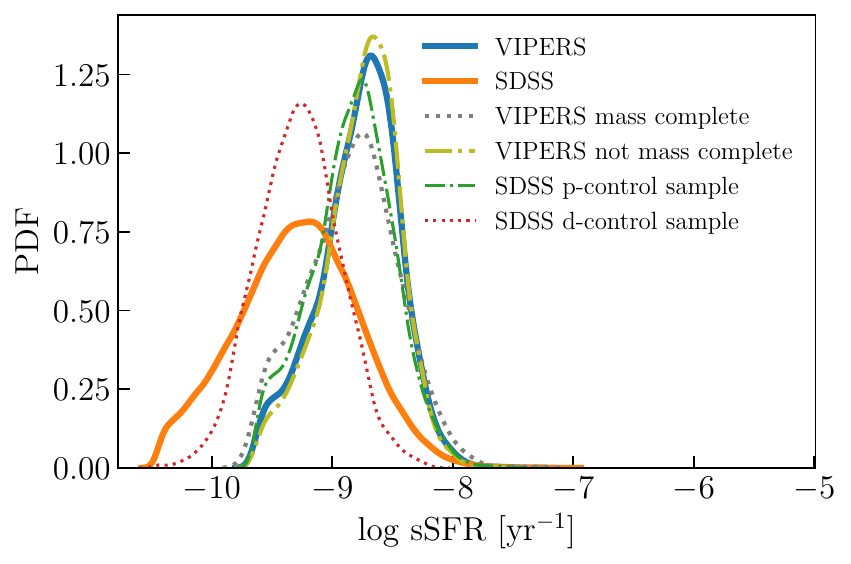}\includegraphics{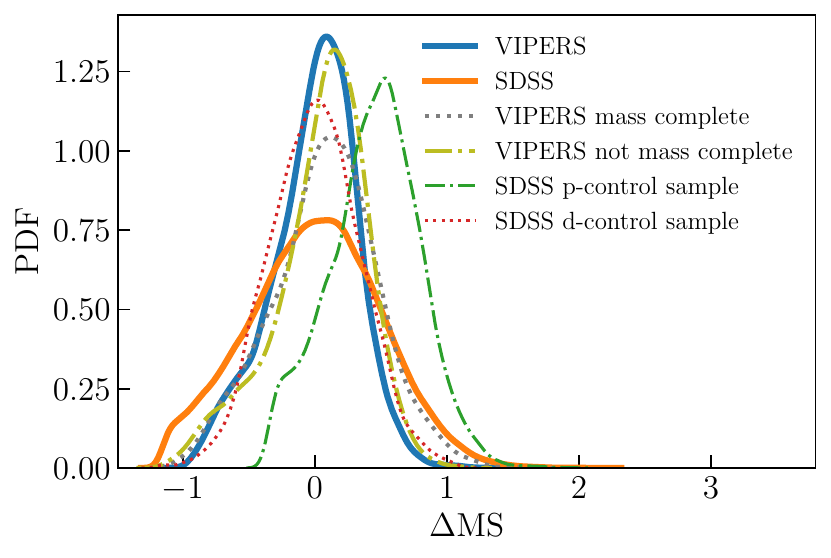}}
    \caption{Kernel density estimations (KDEs) of $M_\star$ (upper left panel), SFR (upper mid panel), metallicity (upper right panel), redshift (bottom left panel), sSFR (bottom mid panel), and the SFR difference with respect to the corresponding MS value (bottom right panel) for the VIPERS (blue solid line), VIPERS mass-complete (gray dotted line), VIPERS not mass-complete (olive dash-dotted line), SDSS (orange solid line), SDSS p-control (green dash-dotted line), and SDSS d-control (red dotted line) samples.}
    \label{fig:dists}
\end{figure*}

Figure~\ref{fig:dists} presents the kernel density estimation (KDE) of the distributions for the main physical features of the two main samples (VIPERS and SDSS), the VIPERS mass-complete \citep[following][]{davidzon2016env} and what is left out, called VIPERS not mass-complete, and the two control samples (p-control and d-control samples).
This figure shows the distribution of $M_{\star}$, SFR, metallicity, redshift, sSFR, and the distance from the MS.

The VIPERS mass-complete results in having a distribution peaked at higher values for $M_\star$, SFR, and metallicity with respect to the VIPERS sample. The distribution in sSFR results having an excess of more passive galaxies with respect to the VIPERS sample.
The VIPERS not mass-complete, instead, results in having a distribution peaked at slightly lower values for $M_\star$, SFR, and metallicity with respect to the VIPERS sample.

The p-control sample shows a shift towards higher mass with respect to the VIPERS sample; while the SFR distribution shows a good agreement with a small excess at smaller SFR with respect to the VIPERS sample.
These differences are due to the fact that during the building of the p-control sample galaxies with high $M_\star$ and low SFR are selected preferentially due to the higher sampling in SDSS at the bottom-right corner of the $M_\star$-SFR bin because of the shift between the MSs from SDSS to VIPERS samples.
The distribution in metallicity shows a shift between p-control and VIPERS samples with the p-control sample being metal-richer than the VIPERS sample.
By construction, the p-control sample shows a higher sSFR and a bigger distance from the MS with respect to the SDSS sample.
Again, the distribution in sSFR has an excess at lower sSFR with respect to the VIPERS sample.

The d-control sample instead recovers the distribution in $M_\star$.
The distributions in SFR, metallicity, and sSFR show a shift toward lower values with respect to the SDSS sample.
Both control samples have a reduced redshift range showing a narrower distribution with respect to the SDSS sample, and the construction of the control samples removed the high-$z$ end of the SDSS distribution.

Table~\ref{tab:diff_control} summarizes the differences of the median values of $M_\star$, SFR, and sSFR for all control samples with respect to VIPERS. 
The distributions of the shift from the MS (defined as the SFR-difference between the measured value and the MS value, shown in the bottom right panel of Fig.~\ref{fig:dists}) also differ with the d-control sample having a much narrower distribution than the other samples.
\begin{table}
\caption{Differences of the 
median values of $M_\star$, SFR, and sSFR between VIPERS and control samples.}              
\label{tab:diff_control}      
\centering                                      
\begin{tabular*}{\columnwidth}{l @{\extracolsep{\fill}} c  c}          
\hline    
\hline
\noalign{\smallskip}
Difference & p-control &  d-control \\    
\hline                                   
\noalign{\smallskip}
    $\Delta \log M_\star$ & $- 0.1 \pm 0.5$ & $- 0.0 \pm 0.5$\\      
    $\Delta \log \text{SFR}$ & $- 0.0 \pm 0.6$ & $+ 0.5 \pm 0.6$\\
    $\Delta \log \text{sSFR}$ & $+ 0.1 \pm 0.5$ & $+ 0.5 \pm 0.5$ \\
    \noalign{\smallskip}
    \hline
\end{tabular*}
\end{table}


\section{Methods of comparison}\label{sec:methods}

In order to compare samples at different redshifts (from $z \sim 0.8$, VIPERS, to $z \sim 0$, SDSS), we apply two families of methods to study the FMR: i) parametric: a) the study of the projections of the FMR; b) metallicity difference in $M_\star$-SFR bin ii) non-parametric: the study of the normalized metallicity-sSFR relation in different mass bins.
The idea for the parametric method is to infer information about the FMR via the median projections on different planes.
Here, we study: i) the MZR \citep{tremonti2004origin, savaglio2005gemini, mannucci2010fundamental, curti2020mass}; ii) the metallicity-SFR relation; iii) the metallicity-sSFR relation; iv) the projection of minimum scatter \citep{mannucci2010fundamental}. 
The study of the projections of the FMR has the problem of being affected by biases introduced by the observations and by the data selection \citep{pistis2022bias}, especially the MZR and the metallicity-SFR relation.
In this method, it is necessary to cross-match the samples in order to compare galaxies with specific properties.
This, in turn, tends to introduce additional selection effects to the samples.
A second parametric method infers information by studying the metallicity difference between samples in $M_\star$-SFR bins projected over the MS.
This method compares directly galaxies with similar physical properties.

The non-parametric method \citep{salim2014critical, salim2015mass} studies the relation between the normalized sSFR and the metallicity.
Because of the use of the sSFR, this method is independent of the simple shift of $M_\star$ and/or SFR resulting from different techniques of estimation of those physical values, e.g., if different samples have physical properties derived using different IMF, the sSFR takes into account automatically the shift.
This method has the advantage of using the projection of the FMR that is less affected by biases \citep{pistis2022bias}.
The original method \citep{salim2014critical, salim2015mass} defines the normalized sSFR as
\begin{equation}
    \Delta \log \text{sSFR} = \log \text{sSFR} - \left< \log \text{sSFR} \right>,
\end{equation}
where $\left< \log \text{sSFR} \right>$ is the average or median sSFR in the mass bin of the sample at low redshift, for both samples.
We also divide the sample into four $M_\star$ bins centered at $\log M_\star \left[ M_\sun \right] =9.5$, $10.0$, $10.5$, and $11.0$ with bin width equal to $0.5$ dex.
This normalization allows comparing galaxies with the same $M_\star$ and SFR.

In this paper, we decide to use a second normalization defined as
\begin{equation}
    \delta \log \text{sSFR} = \log \text{sSFR} - \log \text{sSFR}_\text{MS},
\end{equation}
where $\log \text{sSFR}_\text{MS}$ is the ``local'' MS at the redshift of each sample.
This normalization allows comparing galaxies with the same relative distance from the MS, giving information about the processes that generate the scatter of the MS itself.

The anti-correlation between metallicity and sSFR has been interpreted in terms of gas accretion from the IGM and circumgalactic medium 
\citep[CGM,][]{mannucci2010fundamental, curti2020mass, kumari2021fmr}.
The accreted gas dilutes the gas metallicity and enhances the star formation.
However, the large scatter of sSFR per fixed metallicity means that galaxies with the most significant offsets from the MS are not always those with fewer metals with respect to their MS counterparts.
This implies the complexity of physical mechanisms in galaxy evolution, e.g., environmental effects \citep[shock heated gas in overdensities cannot cool down efficiently and galaxies become metal-rich rapidly due to the suppression of pristine gas inflow,][]{lilly2013fmr, peng2014evo}.

Galaxies with negative scatter ($\delta \log \text{sSFR} < 0$) with respect to the MS were most probably undergo a quenching process in the recent history \citep[e.g., via depletion or outflows of gas;][]{ciesla2017quenching, ciesla2018drop}, while galaxies with a positive scatter ($\delta \log \text{sSFR} > 0$) with respect to the MS experienced a recent SFR enhancement \citep[burst, e.g., via merging event or inflows of gas;][]{elbaz2018burst}.

In order to account for these processes, we 
divide the samples into the sub-samples of galaxies above the MS ($\delta \log \text{sSFR} > 0$) and below the MS ($\delta \log \text{sSFR} < 0$).
By studying the slope of the normalized (according to the MS) metallicity-sSFR relation as a function of the mass, we can infer the impact of processes that enhance or quench the SFR on the metallicity during galaxy evolution.

In the following sections (Sect.~\ref{sec:comparison} and Sect.~\ref{sect:mzrevo}), we study the median relation on different planes, following the binning scheme described in Table~\ref{tab:bin}.
\begin{table*}
\caption{Binning scheme for VIPERS and SDSS (main, p-control, and d-control) samples according to the property to bin.}              
\label{tab:bin}      
\centering                                      
\begin{tabular*}{\textwidth}{l @{\extracolsep{\fill}} c c c c}          
\hline
\hline
 \noalign{\smallskip}
 & \multicolumn{2}{c}{VIPERS-based samples and} & \multicolumn{2}{c}{SDSS main sample}\\
  & \multicolumn{2}{c}{SDSS-based control samples} & \multicolumn{2}{c}{}\\
   \cline{2-3}  \cline{4-5}\\
\multicolumn{1}{l}{Property (to bin)} & \multicolumn{1}{c}{Range} & \multicolumn{1}{c}{Method} & \multicolumn{1}{c}{Range} & \multicolumn{1}{c}{Method}\\
   \noalign{\smallskip}
\hline                                   
 \noalign{\smallskip}
 $\log M_\star \left[ M_\sun \right]$ & $\left[ 8.43 , 11.47 \right]$ & Bin width 0.15 dex & $< 9.0$ & Equal number of galaxies \\
 & & & $\left[ 9.0, 11.5 \right]$ & Bin width 0.15 dex \\
  & & & $> 11.5$ & Equal number of galaxies \\
  ${\log \text{SFR}  \left[ M_\odot \text{ yr}^{-1} \right]}$ & $\left[ -0.9, -1.5 \right]$ & Bin width 0.15 dex & ${< -1.0}$ & Equal number of galaxies \\
  & & & ${< -1.0}$ & Bin width 0.15 dex \\  
$\log \text{sSFR} \left[ \text{yr}^{-1} \right]$ & $\left[ -11.8, -7.8 \right]$ & Bin width 0.15 dex & $\left[ -12.0, -7.5 \right]$ & Bin width 0.15 dex \\
     \noalign{\smallskip}
\hline  
\end{tabular*}
\end{table*}
For a fair comparison with the literature, we use the same bin width used in \cite{curti2020mass} when binning with a constant bin width.
We used a different method of binning the properties of SDSS in the range covered by this sample but not by the VIPERS sample.
We keep only bins with a number of galaxies higher than 25 \citep[following][]{curti2020mass}.

We defined two errors based on: i) the distribution of the population inside the bin $\sigma_\mathrm{dist}$ from the $84$th and $16$th percentile (equivalent to $68\%$ of the population inside each bin), and ii) the error on the median $\sigma_\mathrm{med} = \sigma_\mathrm{dist}/\sqrt{N}$ ($N$ is the number of galaxies in the bin). In the following analysis, we use $\sigma_\mathrm{med}$ to draw conclusions about the possible evolution of the considered relations, while $\sigma_\mathrm{dist}$ provides us with information about their scatter. 

\section{Comparison of FMR between different redshift ranges}\label{sec:comparison}

In this section, we present the results of the comparison of the FMR of the main samples at median ${z \sim 0.09}$ (SDSS), median ${z \sim 0.63}$ (VIPERS), and the control samples.
In Sect.~\ref{sec:cross} we report the results using the parametric method.
In Sect.~\ref{sec:nonparam} we report the results using the non-parametric method.


\subsection{Parametric method I: FMR projections with control samples}\label{sec:cross}

\begin{figure*}
    \centering
    \resizebox{\hsize}{!}{\includegraphics{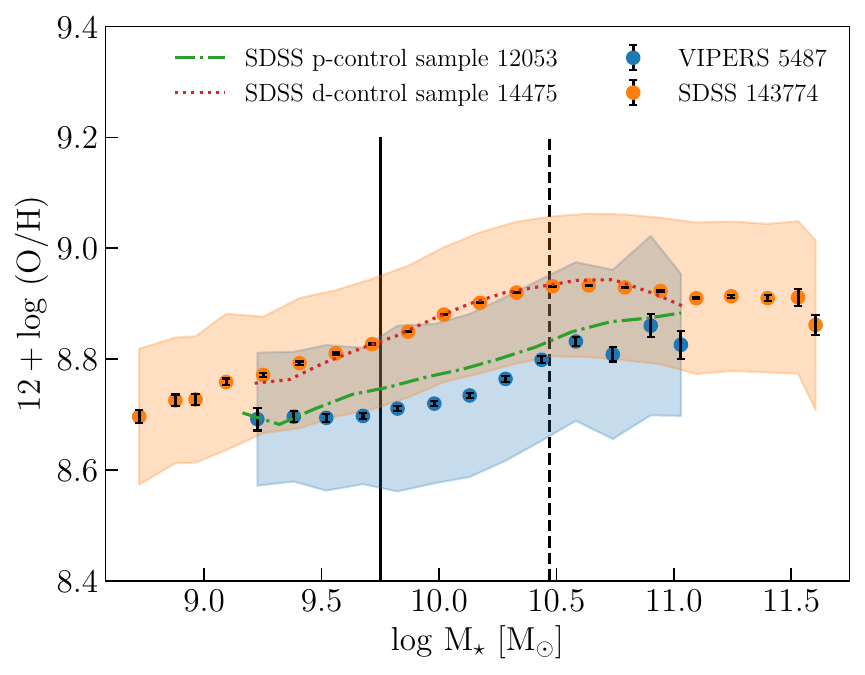}\includegraphics{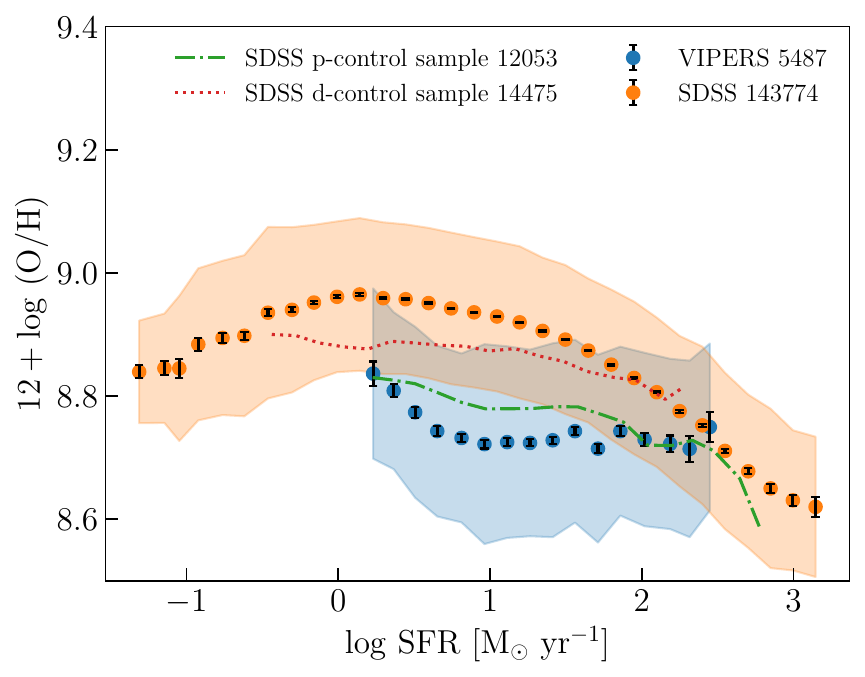}}
    \resizebox{\hsize}{!}{\includegraphics{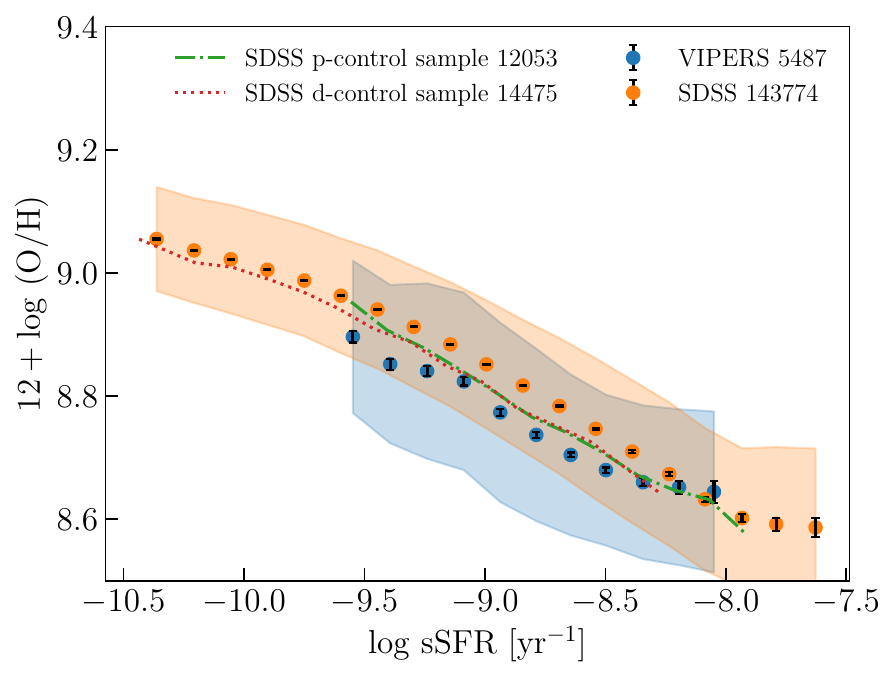}\includegraphics{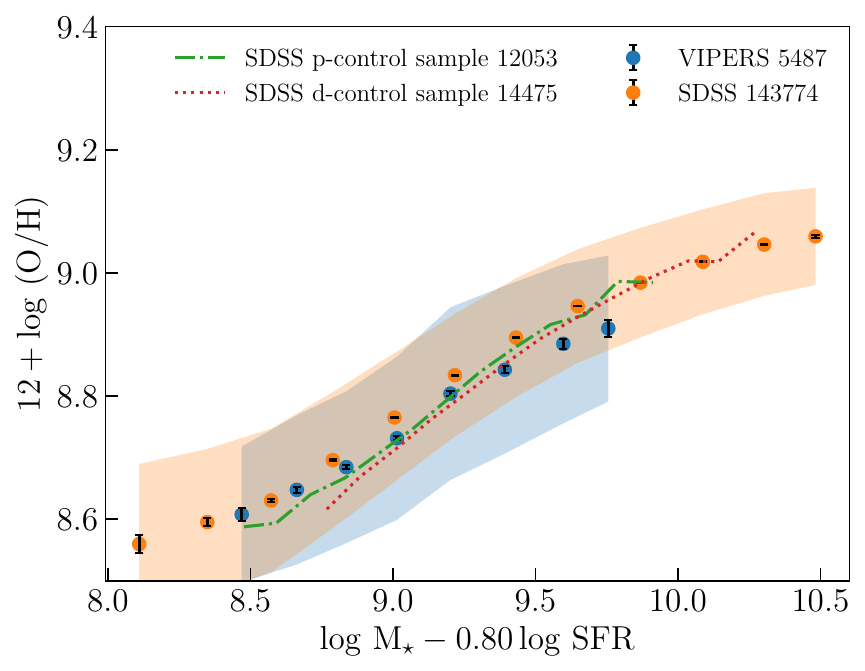}\llap{\shortstack{%
        \includegraphics[scale=.3]{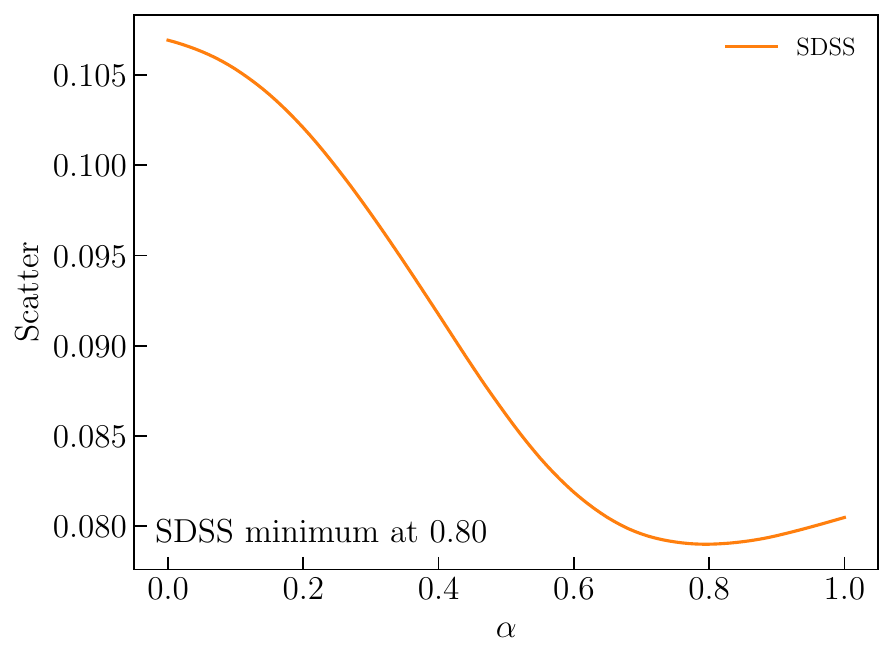}\\
        \rule{0ex}{2.25in}%
      }
  \rule{3in}{0ex}}}
    \caption{Four projections of the FMR: MZR (top left),metallicity-SFR relation (top right), metallicity versus. sSFR (bottom left), and metallicity versus. ${\log M_\star - \alpha \log \text{SFR}}$ (bottom right, inside this panel, the scatter as a function of $\alpha$ is reported) for VIPERS (blue dots), SDSS (orange dots), SDSS p-control (green dash-dotted line), and SDSS d-control (red dotted line). The shaded areas show the $1 \sigma_\mathrm{dist}$ while the black errorbars show the $1 \sigma_\mathrm{med}$ for the metallicity. We report the number of galaxies in each sample in the legend. The vertical black solid line indicates the $M_\star$ (${\log M_\star \left[ M_\sun \right] = 9.75}$) below which the MZR at intermediate redshift flattens. The vertical black dashed line in the MZR plane shows the most conservative mass limit for completeness (${\log M_\star \left[ M_\sun \right] = 10.47}$ for ${0.65 < z <= 0.8}$) in the redshift range observed by VIPERS \citep{davidzon2016env}. For each sample, we report the number of galaxies in the legend.}
    \label{fig:fmr_proj}
\end{figure*}

We proceed with the comparison of the samples at median ${z \sim 0.09}$ (SDSS), median ${z \sim 0.63}$ (VIPERS), and the SDSS-based control samples via the parametric method.
Figure~\ref{fig:fmr_proj} shows the projections of the FMR on the $M_\star$, SFR, sSFR, and the plane of minimum scatter $\mu_\alpha = \log M_\star - \alpha \log \text{SFR}$ \citep{mannucci2010fundamental} planes. 

\subsubsection{MZR}

The MZR (Fig.~\ref{fig:fmr_proj}, upper left panel) of galaxies at intermediate redshift shows lower metallicities at a given $M_\star$ with respect to galaxies at low redshift.
This shift is statistically significant with respect to $\sigma_\mathrm{med}$ showing an evolution of the MZR with the redshift.

However, a large scatter of both populations should be noted.
As seen from the upper left panel of Fig.~\ref{fig:fmr_proj}, the separation between them is below $1 \sigma_\mathrm{dist}$ of both samples.
The samples at both low and intermediate redshifts show a similar scatter, given by $\sigma_\mathrm{dist}$, suggesting a lack of evolution of the scatter itself.
This implies that the detection or non-detection of the evolution of MZR may be sensitive to source selection in small samples.

The p-control sample shows a similar MZR as the VIPERS sample, while the d-control sample follows the same MZR as the SDSS sample.
However, the p-control sample does not show a strong flattening as the VIPERS sample at low $M_\star$ (${\log M_\star \left[ M_\sun \right] \lesssim 9.75}$).
This flattening could be due to the lack of observation/selection of high-SF galaxies at low $M_\star$, due to the VIPERS sample being mass-complete only above ${\log M_\star \left[ M_\sun \right] \gtrsim 10.50}$.
The same flattening is observed in VVDS sample \citep{lamareille2009vimos, perez2009mzr} which has the same survey depth as VIPERS.
This counter-intuitive bias which results in a flattening at low $M_\star$ of the MZR needs to be explored in more detail.



\subsubsection{Metallicity-SFR relation}

The metallicity-SFR relation (Fig.~\ref{fig:fmr_proj}, upper right panel) of galaxies at intermediate redshift shows lower metallicities at a given SFR with respect to galaxies at low redshift.
Like the case of the MZR, the metallicity-SFR relation shows a shift statistically significant with respect to $\sigma_\mathrm{med}$, showing an evolution of the relation with the redshift.
The difference between low and intermediate redshift decreases at high SFR.

Again, a large scatter of both populations should be noted.
As seen from the upper right panel of Fig.~\ref{fig:fmr_proj}, the separation between them is below $1 \sigma_\mathrm{dist}$ of both samples.
The samples at both low and intermediate redshifts show a similar scatter, given by $\sigma_\mathrm{dist}$, suggesting a lack of evolution of the scatter itself.

The p-control shows a closer metallicity-SFR relation to the VIPERS sample.
The process of cross-matching removes metal-rich galaxies in the SFR interval of the VIPERS sample when the control samples are built.
However, the p-control sample does not get as close as the VIPERS sample to the SDSS sample at high SFR.
At the same time, the d-control sample shows a weaker correlation between metallicity and SFR with respect to the SDSS sample.
We report in Appendix~\ref{app:sfr} the effects due to different calibrations for SFRs.

\subsubsection{Metallicity-sSFR relation}

The metallicity-sSFR relation (Fig.~\ref{fig:fmr_proj}, bottom left panel) shows smaller differences between low and intermediate redshifts.
The difference between SDSS and VIPERS samples decreases at high sSFR with vipers having a higher metallicity at a given sSFR.
Again, this difference is significant with respect to $\sigma_\mathrm{med}$, showing an evolution of the metallicity-sSFR relation.

Again, a large scatter of both populations should be noted.
As seen from the bottom left panel of Fig.~\ref{fig:fmr_proj}, the separation between them is below $1 \sigma_\mathrm{dist}$ of both samples.
The samples at both low and intermediate redshifts show a similar scatter, given by $\sigma_\mathrm{dist}$, suggesting a lack of evolution of the scatter itself.

All control samples follow the same metallicity-sSFR relation as the SDSS sample with a small shift toward lower metallicity around $\log \mathrm{sSFR} \left[ \mathrm{yr}^{-1} \right] =9$.
The differences between the SDSS and all control samples are negligible in the metallicity-sSFR relation with respect to $\sigma_\mathrm{dist}$.

\subsubsection{Projection of minimum scatter}

For the projection of minimum scatter \citep[defined in][]{mannucci2010fundamental}, the $\alpha$ value used to define the ${\mu_\alpha = \log M_\star - \alpha \log \text{SFR}}$ depends on the calibrations used \citep[different calibrations have different values for $\alpha$][]{andrews2013mass, sanders2021mosdefevo}.
For our set of properties, we find \citep[following][]{sanders2021mosdefevo} $\alpha = 0.72$ (Fig.~\ref{fig:fmr_proj}, inside bottom right panel).

The $\mu_\alpha$-metallicity plane (Fig.~\ref{fig:fmr_proj}, bottom right panel), shows an increasing difference at lower values of $\mu_\alpha$.
On this projection of the FMR, both control samples follow the same relation as the SDSS sample.

The projection of minimum scatter shows smaller differences between low and intermediate redshifts.
Again, this difference is significant with respect to $\sigma_\mathrm{med}$, especially at high $\mu_\alpha$ values.

Again, a large scatter of both populations should be noted.
As seen from the bottom right panel of Fig.~\ref{fig:fmr_proj}, the separation between them is below $1 \sigma_\mathrm{dist}$ of both samples.
The samples at both low and intermediate redshifts show a similar scatter, given by $\sigma_\mathrm{dist}$, suggesting a lack of evolution of the scatter itself.

\subsection{Parametric method I: Surface of the fundamental metallicity relation}

\begin{figure}
    \centering
    \resizebox{\hsize}{!}{\includegraphics{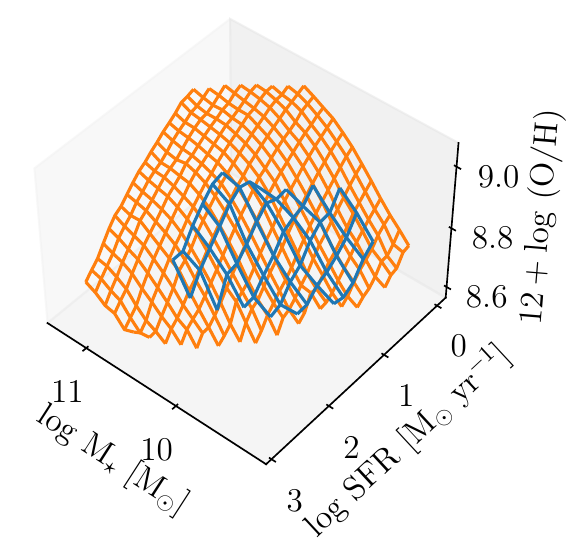}}
    \caption{Surfaces of the FMR for SDSS (orange) and VIPERS (blue) samples.}
    \label{fig:3dfmr}
\end{figure}

Figure~\ref{fig:3dfmr} shows the surfaces of the FMR for both SDSS and VIPERS samples.
The shapes of the two surfaces agree with each other at high $M_\star$.

\subsubsection{Metallicity difference in $M_\star$-SFR bins}

\begin{figure*}
    \centering
    \resizebox{\hsize}{!}{\includegraphics{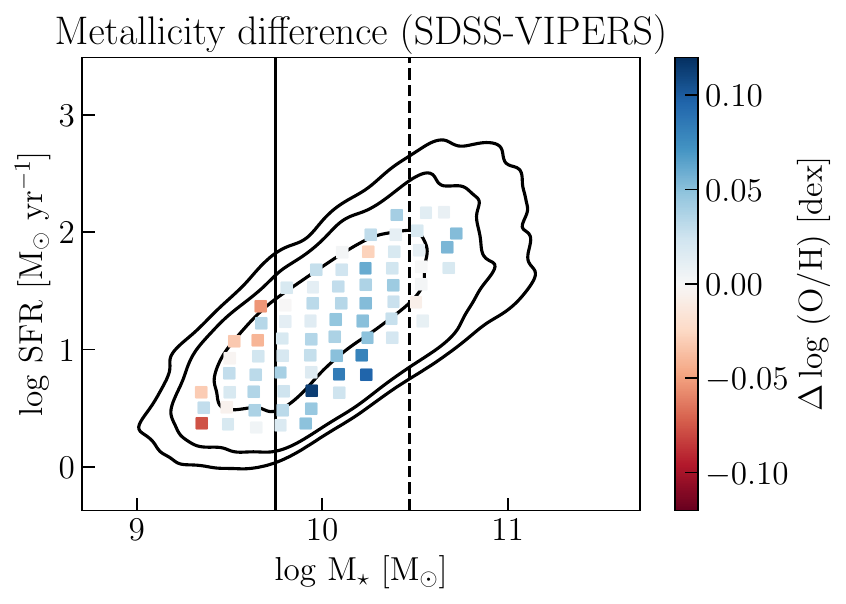}\includegraphics{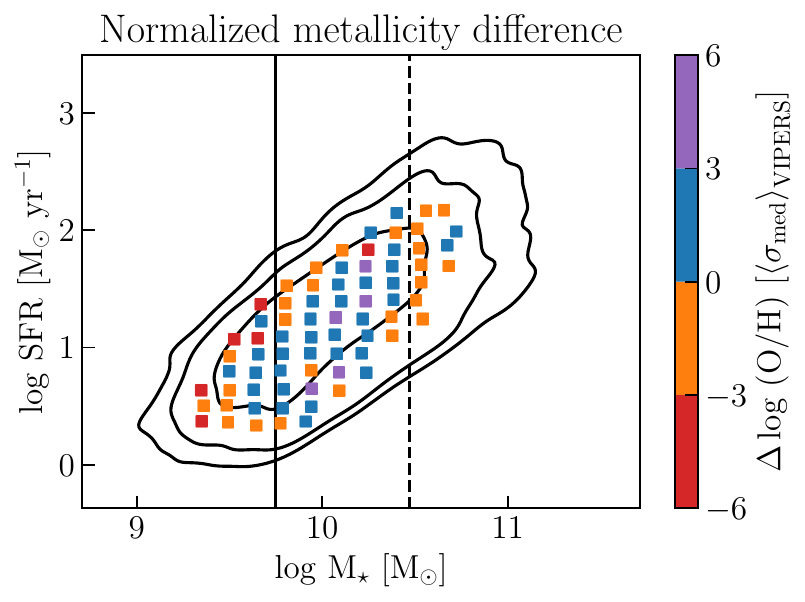}}
    \caption{Metallicity differences $\Delta \log \left( \text{O}/\text{H} \right)$ between SDSS and VIPERS samples. Contours show the $1\sigma$, $2\sigma$, and $3\sigma$ levels of the MS distributions for the VIPERS sample. The vertical black solid line indicates the mass below which the MZR at intermediate redshift flattens. The vertical black dashed line shows the most constrictive mass limit for completeness (${\log M_\star \left[ M_\sun \right] = 10.47}$ for ${0.65 < z <= 0.8}$) in the redshift range observed by VIPERS \citep{davidzon2016env}. In the left panel, the $M_\star$-SFR bins are color-coded according to the metallicity difference in unit of dex. In the right panel, the bins are color-coded according to the metallicity difference normalized by the $\left < \sigma_\mathrm{med} \right>_\mathrm{VIPERS}$.}
    \label{fig:delta_fmr}
\end{figure*}

Figure~\ref{fig:delta_fmr} presents the metallicity difference between SDSS and VIPERS samples for each $M_\star$-SFR bin.
This is the most direct comparison of the FMR between samples.

The average difference in metallicity between all SDSS-based and VIPERS samples is listed in Table~\ref{tab:fmr_diff} where $\left< \sigma \right>_\mathrm{VIPERS}$ is the average standard deviation in metallicity inside the $M_\star$-SFR bin of VIPERS sample.
\begin{table*}
\caption{Average (over bins) differences in metallicity between all SDSS-based and VIPERS samples (as in Fig.~\ref{fig:delta_fmr}) in $M_\star$-SFR bins. The differences are expressed in absolute units, in units of $\left < \sigma_\mathrm{dist} \right>_\mathrm{VIPERS}$, and in units of $\left < \sigma_\mathrm{med} \right>_\mathrm{VIPERS}$.}              
\label{tab:fmr_diff}      
\centering                                      
\begin{tabular*}{\textwidth}{l @{\extracolsep{\fill}} c c c}          
\hline    
\hline
\noalign{\smallskip}
\multicolumn{1}{c}{Sample} & \multicolumn{1}{c}{$\Delta \log \left( \text{O}/\text{H} \right)$} & \multicolumn{1}{c}{$\Delta \log \left( \text{O}/\text{H} \right)$} & \multicolumn{1}{c}{$\Delta \log \left( \text{O}/\text{H} \right)$} \\    
 &  $\left( \left< \sigma_\mathrm{dist} \right>_\mathrm{VIPERS} \right)$ & $\left( \left< \sigma_\mathrm{med} \right>_\mathrm{VIPERS} \right)$ & (dex) \\
\hline                                   
\noalign{\smallskip}
    SDSS main sample & $0.223$ & $1.783$ & $0.028$\\      
    SDSS p-control sample & $0.201$ & $1.611$ & $0.026$\\
    SDSS d-control sample & $0.224$ & $1.931$ & $0.027$ \\
    \noalign{\smallskip}
    \hline
\end{tabular*}
\tablefoot{The difference found in this work is about half of the difference found in \cite{pistis2022bias}. This change is due to the different catalog of line measurements used.}
\end{table*}
The small changes in the $\Delta \log \left( \mathrm{O}/ \mathrm{H} \right)$ values for the control samples are mainly due to the limited area in the MS observed in these sub-samples with respect to the main sample.

We find an average metallicity difference ${\Delta \log \left( \text{O}/\text{H} \right) = 1.78 \left< \sigma_\mathrm{med} \right>_\mathrm{VIPERS}}$.
The majority of bins have a metallicity difference ${\Delta \log \left( \text{O}/\text{H} \right)}$ between ${\pm 3 \sigma_\mathrm{med, VIPERS}}$.
If we apply a stronger selection on $\mathrm{H}\beta$ (${\mathrm{S/N} > 5}$), the difference is reduced to ${\Delta \log \left( \text{O}/\text{H} \right) \sim 1.15 \left< \sigma_\mathrm{med} \right>_\mathrm{VIPERS}}$.

\subsection{Non-parametric method}\label{sec:nonparam}


In this subsection, we proceed to compare the main samples at median $z\sim 0.09$ (SDSS) and median $z \sim 0.63$ (VIPERS) via the non-parametric method.
First, we normalize the sSFR of both samples by the median sSFR of the low redshift sample, to assure that we compare galaxies with both the same $M_\star$ and SFR.

\begin{figure*}
    \centering
    \resizebox{\hsize}{!}{\includegraphics{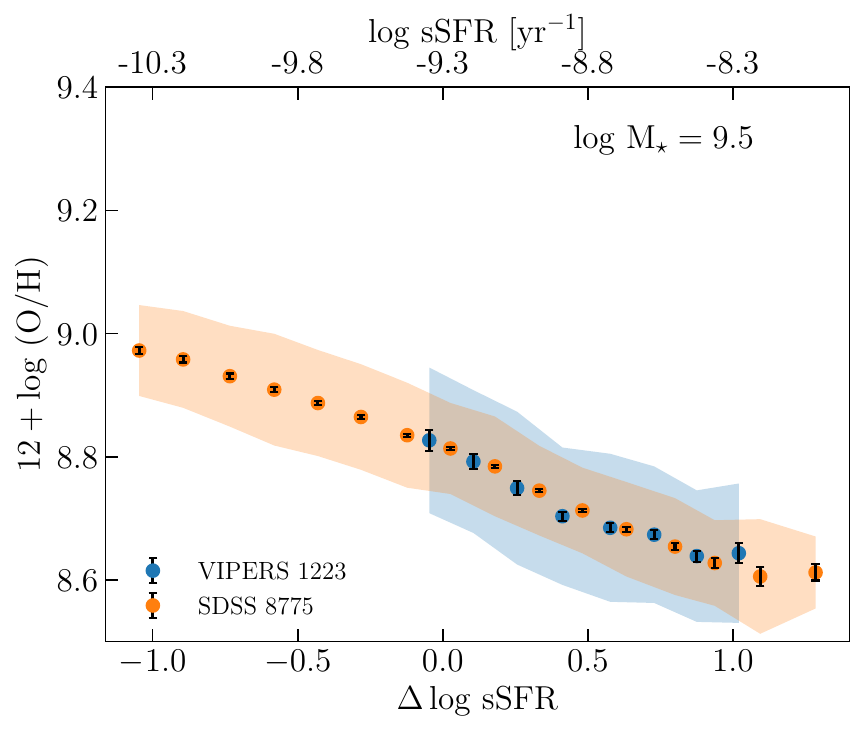}\includegraphics{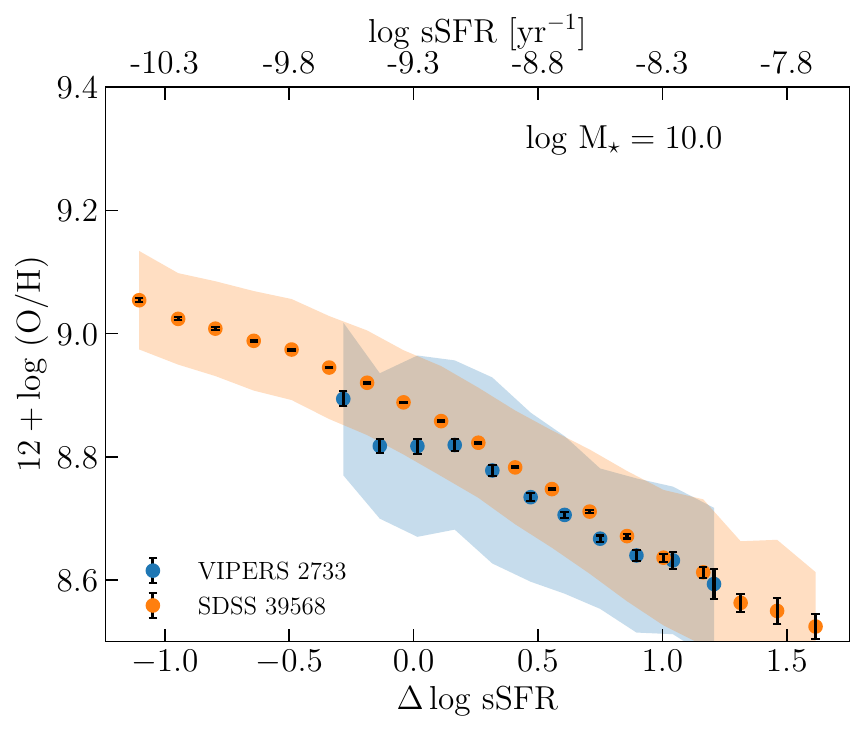}}
    \resizebox{\hsize}{!}{\includegraphics{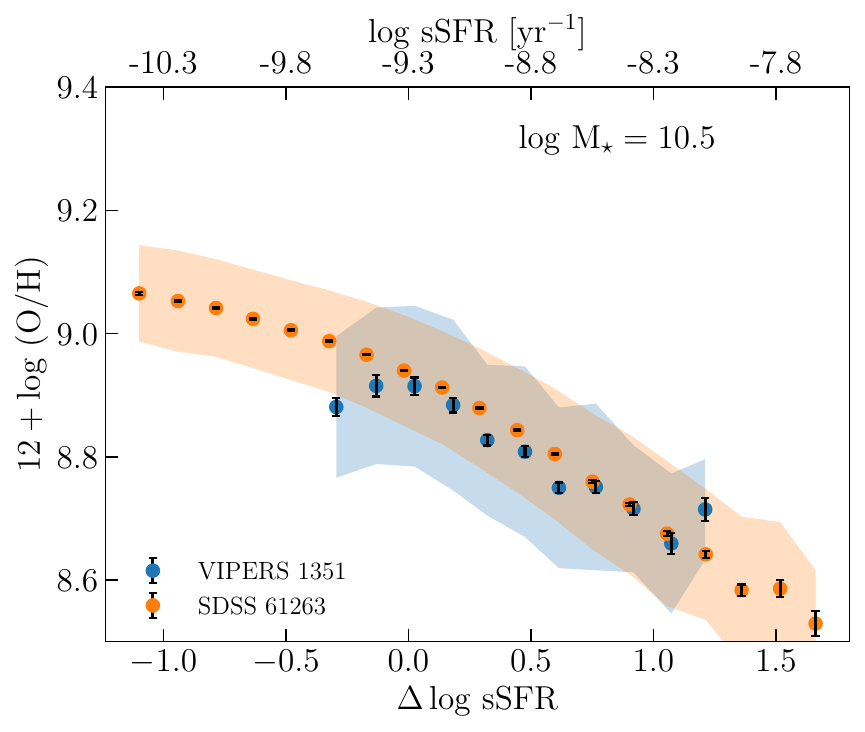}\includegraphics{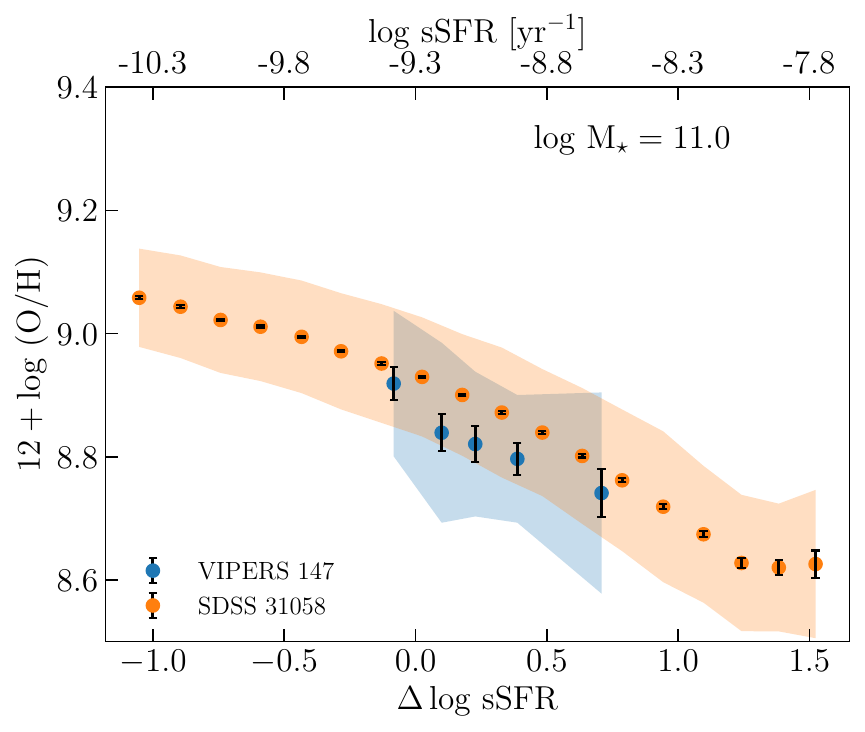}}
    \caption{Comparison in the non-parametric method \citep{salim2014critical, salim2015mass} between VIPERS (blue dots) and SDSS (orange dots) samples. The shaded areas show the $1 \sigma_\mathrm{dist}$ while the black errorbars show the $1 \sigma_\mathrm{med}$ for the metallicity. Mass bins are centered on the values indicated in each panel and are $0.5$ dex wide. We also report the number of galaxies for both samples in each mass bin.}
    \label{fig:comp_fmr_salim}
\end{figure*}

Inside each mass bin, we divided the sample in $0.15$ dex-wide bins in $\Delta \log \text{sSFR}$.
Then, we estimated the median in each bin and we estimate the errors according to $\sigma_\mathrm{dist}$ and $\sigma_\mathrm{med}$
Figure~\ref{fig:comp_fmr_salim} shows the relations for the main samples.
The difference between the two samples increases with the $M_\star$.
This increasing difference with $M_\star$ is in agreement with the differences in metallicity between SDSS (full, p, and d-control) samples (as shown in Fig.~\ref{fig:delta_fmr}).
The small differences in the low-mass bins can be dominated by the flattening at low-mass in the MZR of VIPERS sample.



\begin{figure*}
    \centering
    \resizebox{\hsize}{!}{\includegraphics{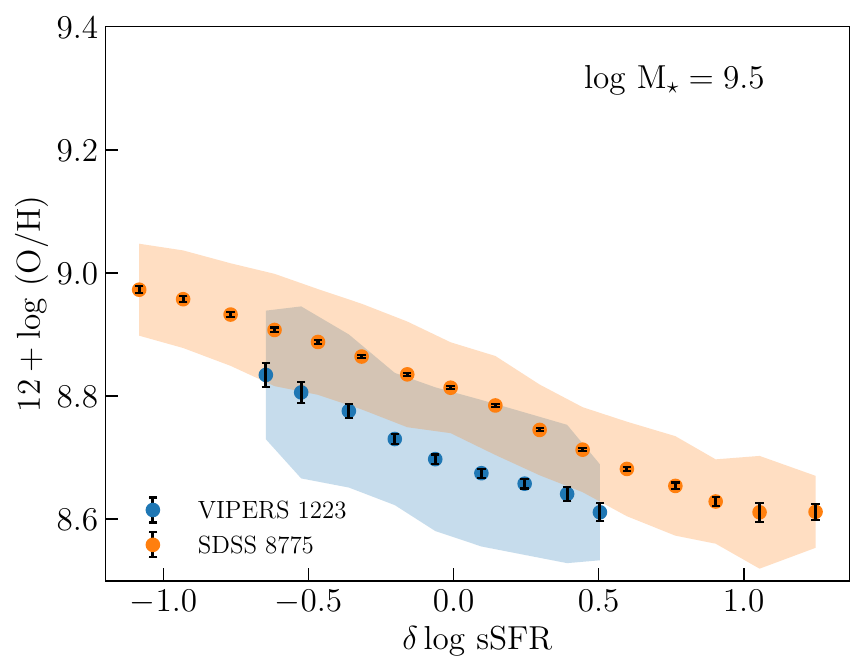}\includegraphics{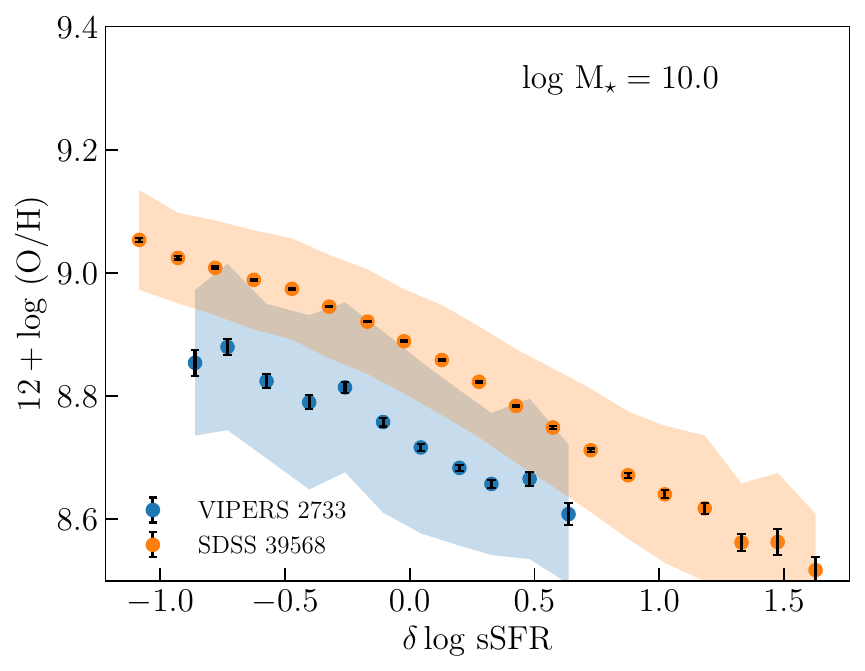}}
\resizebox{\hsize}{!}{\includegraphics{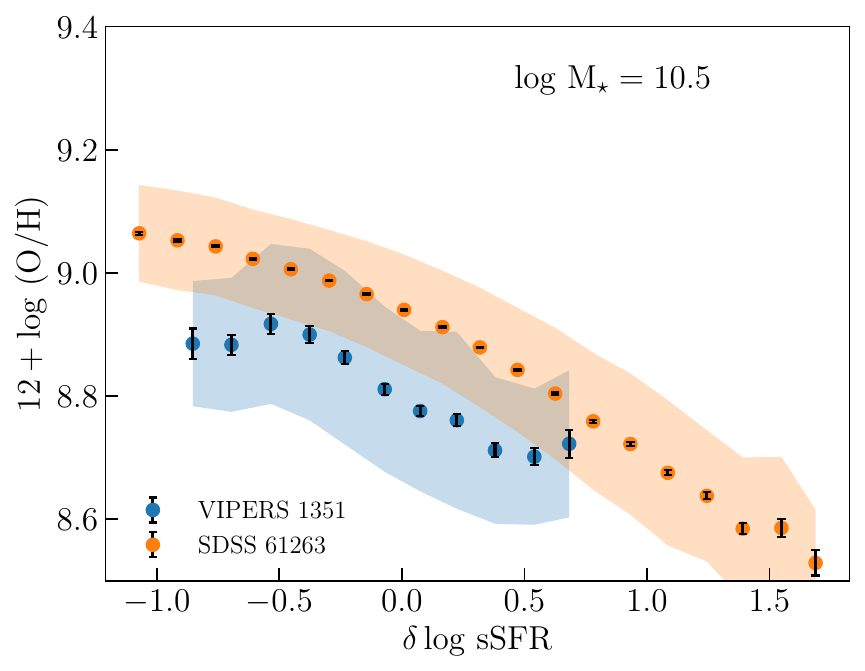}\includegraphics{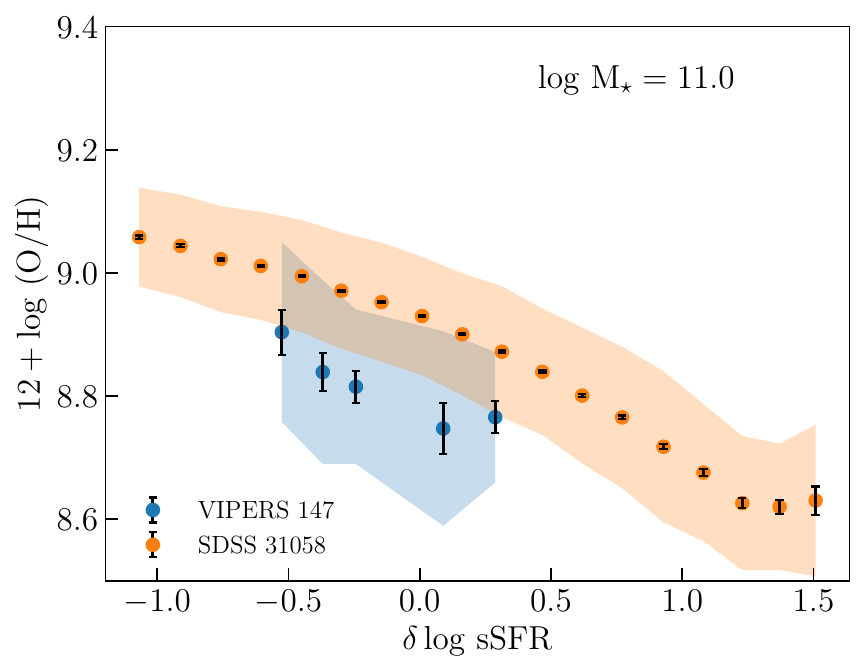}}
    \caption{Comparison in the non-parametric method \citep{salim2014critical, salim2015mass} normalized according to the MS between VIPERS (blue dots) and SDSS (orange dots) samples. The shaded areas show the $1 \sigma_\mathrm{dist}$ while the black errorbars show the $1 \sigma_\mathrm{med}$ for the metallicity. Mass bins are centered on the values indicated in each panel and are $0.5$ dex wide. We also report the number of galaxies for both samples in each mass bin.}
    \label{fig:comp_fmr_relms_ssfr}
\end{figure*}

Then, we normalize the sSFR of each sample by the value of their ``local'' MS.
In this way, we compare galaxies with the same relative distance to the MS.
We proceeded in the same way as the non-parametric method with the division in the same $M_\star$ bins.
In each mass bin, we divided the sample in $0.15$ dex-wide bin in $\delta \log \text{sSFR}$.
Then, we estimated the median in each bin and we estimated the errors according to $\sigma_\mathrm{dist}$ and $\sigma_\mathrm{med}$.
Figure~\ref{fig:comp_fmr_relms_ssfr} shows bigger differences between the samples compared to the normalization with respect to the median value.

\begin{figure}
    \centering
    \resizebox{\hsize}{!}{\includegraphics{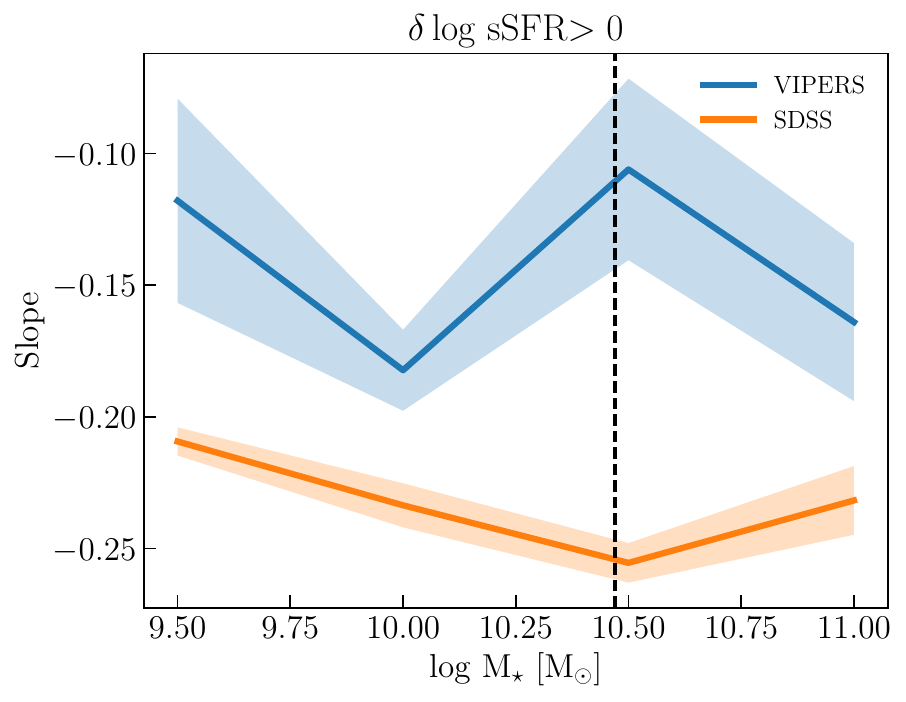}}
    \resizebox{\hsize}{!}{\includegraphics{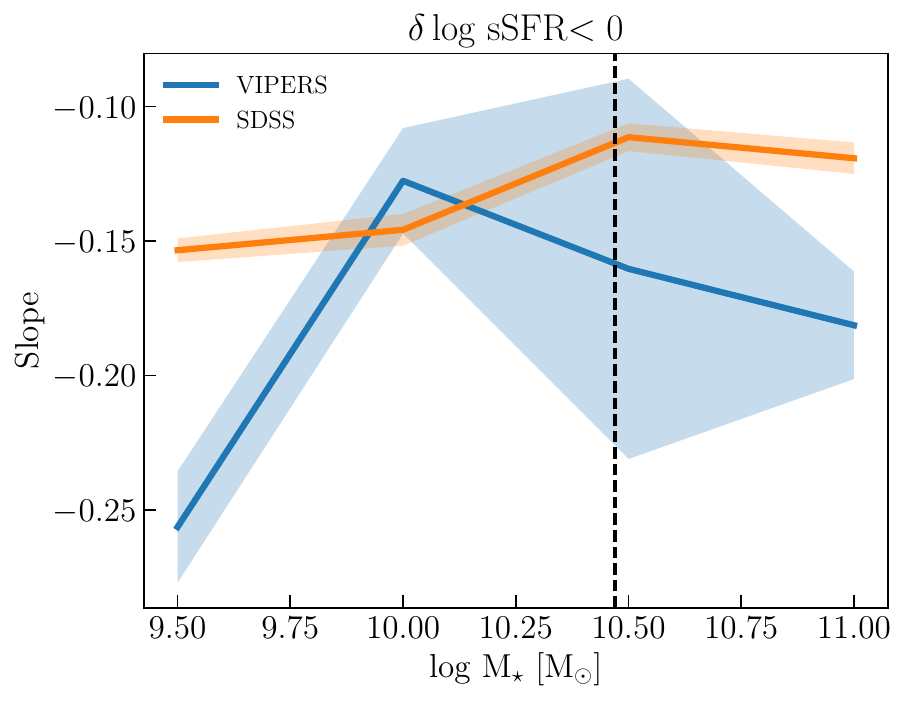}}
    \caption{Slope of the relation between metallicity and $\delta \log \text{sSFR}$ as a function of the $M_\star$ above (top) and below (bottom) the MS for VIPERS (blue solid line) and SDSS (orange solid line) samples. The vertical black dashed line shows the most conservative mass limit for completeness (${\log M_\star \left[ M_\sun \right] = 10.47}$ for ${0.65 < z <= 0.8}$) in the redshift range observed by VIPERS \citep{davidzon2016env}.}
    \label{fig:slope}
\end{figure}

To analyze the processes that lead the galaxies to move around the MS, we study the slope of the normalized metallicity-sSFR relation as a function of the $M_\star$ for galaxies above (${\delta \log \text{sSFR} > 0}$) and below (${\delta \log \text{sSFR} < 0}$) the MS.
Table~\ref{tab:slope} reports the values of the slope.
\begin{table*}
\caption{Slope of the $\delta \log \text{sSFR}$-metallicity relation for $\delta \log \text{sSFR} < 0$ and $\delta \log \text{sSFR} > 0$.}              
\label{tab:slope}      
\centering                                      
\begin{tabular*}{\textwidth}{l @{\extracolsep{\fill}} c c c c}          
\hline                        
\hline
\noalign{\smallskip}
 & \multicolumn{2}{c}{$\delta \log \text{sSFR} < 0$} & \multicolumn{2}{c}{$\delta \log \text{sSFR} > 0$} \\
   \cline{2-3} \cline{4-5}\\
$\log M_\star \left[ M_\odot \right]$ & VIPERS & SDSS & VIPERS & SDSS  \\    
\noalign{\smallskip}
\hline                                   
\noalign{\smallskip}
    $9.5$  & $-0.26 \pm 0.02$ & $-0.153 \pm 0.004$ & $-0.12 \pm 0.04$ & $-0.209 \pm 0.005$\\      
    $10.0$ & $-0.13 \pm 0.02$ & $-0.146 \pm 0.006$ & $-0.18 \pm 0.02$ & $-0.234 \pm 0.008$\\
    $10.5$ & $-0.16 \pm 0.07$ & $-0.111 \pm 0.005$ & $-0.11 \pm 0.03$ & $-0.255 \pm 0.008$\\
    $11.0$ & $-0.18 \pm 0.06$ & $-0.119 \pm 0.006$ & $-0.16 \pm 0.02$ & $-0.232 \pm 0.013$\\
    \noalign{\smallskip}
\hline
\end{tabular*}
\end{table*}
Figure~\ref{fig:slope} shows the slope of the normalized metallicity-sSFR relation as a function of the $M_\star$.
For both samples and both positions with respect to the MS, a negative slope indicates processes that are diluting the metals in the ISM while quenching or enhancing the SFR of the galaxies.

Above the MS ($\delta \log \mathrm{sSFR} > 0$) we find stronger dilution effects from processes enhancing the SFR at low redshift for smaller $M_\star$.
The dilution effects remain almost constant in the whole mass range here explored at low redshift, while more massive galaxies at intermediate redshift have stronger dilution than the less massive galaxies.
The difference between low and intermediate redshifts decreases with $M_\star$.
Below the MS ($\delta \log \mathrm{sSFR} < 0$) we find
a consistent relation within uncertainties between low and intermediate redshifts with less dilution of the metals increasing the $M_\star$.
Again the differences between samples reduce increasing the $M_\star$, with the exception of the last $M_\star$ bin.
Below MS, the dilution of metals remains stronger at intermediate redshift in comparison to the low redshift for the whole range of $M_\star$.


\section{Evolution of the MZR, metallicity-SFR relation, and FMR}\label{sect:mzrevo}

Being particularly careful to homogenize the property estimations of both samples, we want to statistically quantify the evolution (within the uncertainties) of the MZR, metallicity-SFR relation up to $z \sim 0.8$ (Fig.~\ref{fig:fmr_proj} upper panels), and FMR.
The systematic trend of the VIPERS sample having lower metallicities than the SDSS at almost all the examined mass ranges is visible but given the scatter of both samples, it is significant only at the level of $\sigma_\mathrm{med}$. 

A possible reason for this apparent weak evolution of the scatter in the MZR and metallicity-SFR relation could be observational biases in the VIPERS sample, in particular, the fact that the sample is not mass-complete, i.e. with increasing $z$ we lose less bright, and consequently less massive galaxies.
To check if this small evolution between SDSS and VIPERS samples results from the VIPERS lower redshift galaxies dominating the MZR and metallicity-SFR relation of the whole sample, we conducted a series of tests, described below.
To check the impact of the mixture of galaxies at different redshifts, we split the VIPERS sample into two redshift bins (with a threshold at the central redshift $z = 0.65$).

\begin{figure}
    \centering
    \resizebox{\hsize}{!}{\includegraphics{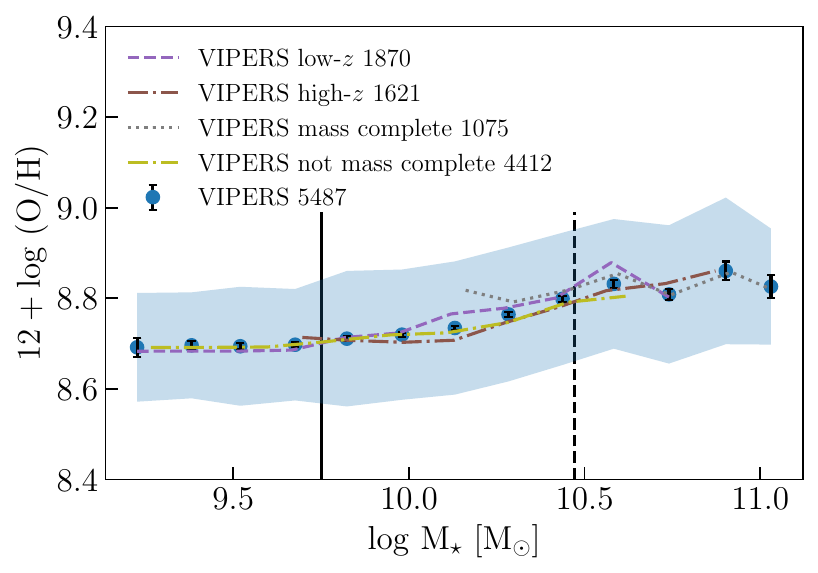}}
    \resizebox{\hsize}{!}{\includegraphics{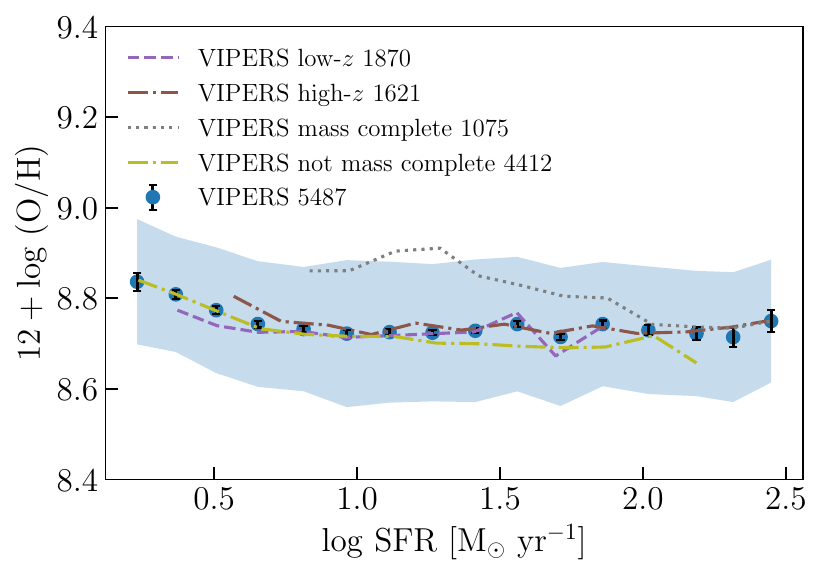}}
    \caption{MZR (top) and the metallicity-SFR relation (bottom) for full VIPERS (blue dots), VIPERS low redshift (purple dashed line), VIPERS high-$z$ (brown dash-dotted line), and VIPERS mass-complete (gray dotted line). The vertical black solid line indicates the mass below which the MZR at intermediate redshift flattens. The vertical black dashed line shows the most constrictive mass limit for completeness (${\log M_\star \left[ M_\sun \right] = 10.47}$ for ${0.65 < z <= 0.8}$) in the redshift range observed by VIPERS \citep{davidzon2016env}. The shaded areas show the $1 \sigma_\mathrm{dist}$ while the black errorbars show the $1 \sigma_\mathrm{med}$ for the metallicity. For each sample, we report the number of galaxies in the legend.}
    \label{fig:evo}
\end{figure}

Figure~\ref{fig:evo} shows the MZR and the metallicity-SFR relation for different VIPERS-based sub-samples.

Any difference in the MZR (upper panel in Fig~\ref{fig:evo}) between the redshift sub-samples is negligible, well below the statistical $\sigma_\mathrm{dist}$ of the measurements.
The MZR for the mass-complete sub-sample results at higher metallicities with respect to the VIPERS sample while the not mass-complete sub-sample follows with better agreement the VIPERS sample.

The metallicity-SFR relation (bottom panel in Fig~\ref{fig:evo}) of the two redshift sub-samples shows bigger differences than the MZR, with the high-$z$ sub-sample having higher metallicity (likely because of being dominated by more massive galaxies) and showing a stronger anti-correlation.
Also, the mass-complete sub-sample shifts towards higher metallicity (also likely because it is dominated by higher mass galaxies) showing a much stronger anti-correlation (see also Appendix~\ref{app:sfr}).
The not mass-complete sub-sample follows the same relation as the VIPERS sample up to ${\log \mathrm{SFR} \left[ M_\sun\ \mathrm{yr}^{-1} \right] \sim 1.25}$ when it shifts toward lower metallicity with respect to the VIPERS sample.
However, all these differences are not statistically significant compared to the $\sigma_\mathrm{dist}$ of the metallicity.

\begin{figure*}
    \centering
    \resizebox{\hsize}{!}{\includegraphics{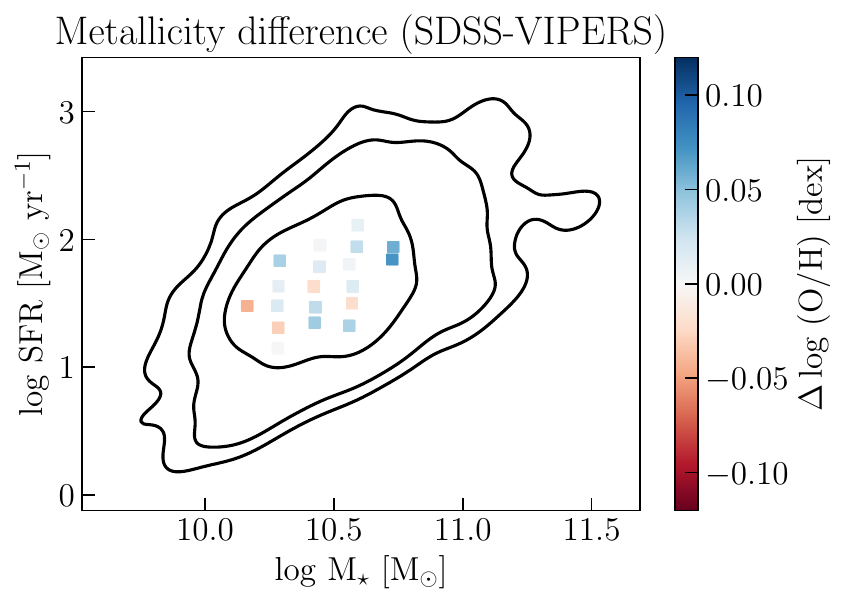}\includegraphics{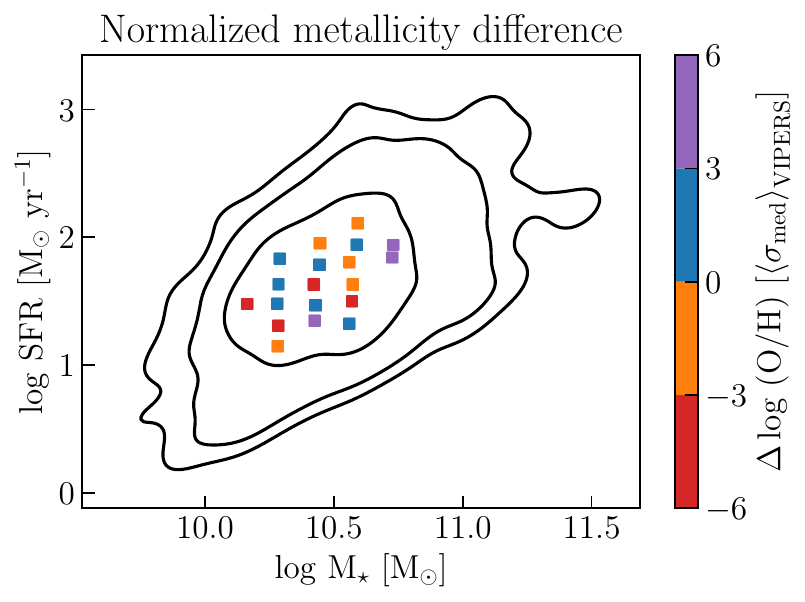}}
    \resizebox{\hsize}{!}{\includegraphics{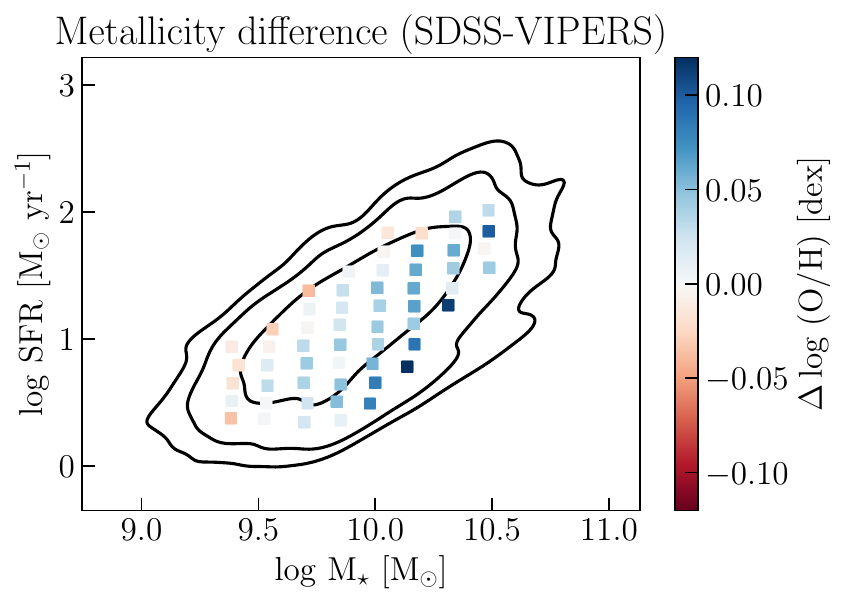}\includegraphics{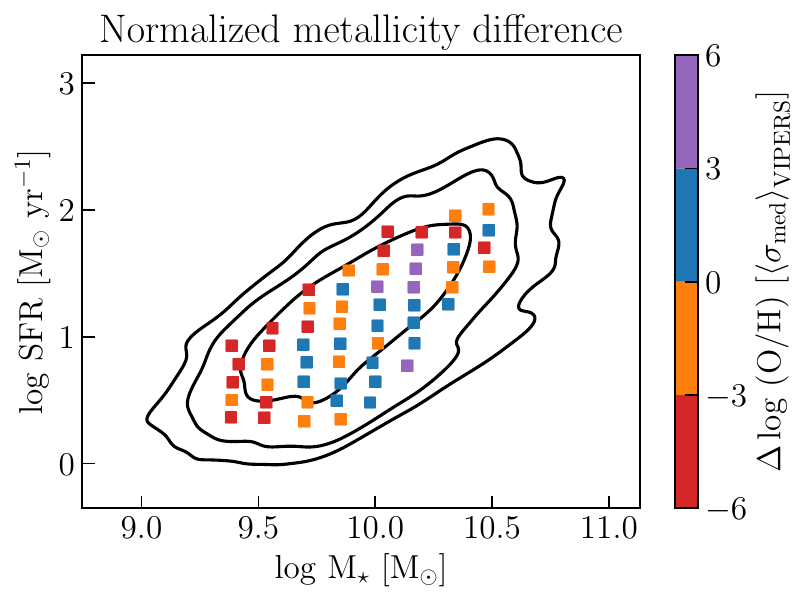}}
    \caption{Metallicity differences $\Delta \log \left( \text{O}/\text{H} \right)$ between SDSS and VIPERS mass-complete samples (top row) and between SDSS and VIPERS not mass-complete samples (bottom row). Contours show the $1\sigma$, $2\sigma$, and $3\sigma$ levels of the MS distributions for the VIPERS sub-samples. In the left panel, the $M_\star$-SFR bins are color-coded according to the metallicity difference in unit of dex. In the right panel, the bins are color-coded according to the metallicity difference normalized by the $\left < \sigma_\mathrm{med} \right>_\mathrm{VIPERS}$.}
    \label{fig:delta_fmr_masscomplete}
\end{figure*}

Regarding the total FMR, the role of a mass-complete sample is the one with the major impact on the results compared to the methodology applied in this study.
Figure~\ref{fig:delta_fmr_masscomplete} shows the metallicity difference between SDSS and (not) mass-complete VIPERS samples.
These are also reported in Table~\ref{tab:fmr_diff_masscomplete} and Table~\ref{tab:fmr_diff_notmasscomplete}.
Once a mass-complete sample is required, the metallicity difference ${\Delta \log \left( \text{O}/\text{H} \right) \left[ \left< \sigma_\mathrm{med} \right>_\mathrm{VIPERS} \right]}$ (Table~\ref{tab:fmr_diff_masscomplete}) is reduced by half with respect to the VIPERS SF sample (Table~\ref{tab:fmr_diff}).
This can be due to the simple fact we are now comparing a much smaller area of the FMR (median values of $\log M_\star \left[ M_\sun \right] = 10.46$, $\log \mathrm{SFR} \left[ M_\sun\ \mathrm{yr}^{-1} \right] = 1.42$, and $12 + \log \mathrm{(O/H)} = 8.83$ and much narrower distributions than the VIPERS sample), removing a large area at low $M_\star$.
The average metallicity difference at low $M_\star$ can be dominated by the observation bias that allows us to see only the brightest low-$M_\star$ galaxies.
However, without taking into account a stellar mass-complete sample, the comparison is not as reliable as once the condition is applied.
\begin{table*}
\caption{Average (over bins) differences in metallicity between all SDSS-based and VIPERS mass-complete samples (as in Fig.~\ref{fig:delta_fmr_masscomplete}) in $M_\star$-SFR bins. The differences are expressed in absolute units, in units of $\left < \sigma_\mathrm{dist} \right>_\mathrm{VIPERS}$, and in units of $\left < \sigma_\mathrm{med} \right>_\mathrm{VIPERS}$.}              
\label{tab:fmr_diff_masscomplete}      
\centering                                      
\begin{tabular*}{\textwidth}{l @{\extracolsep{\fill}} c c c}          
\hline    
\hline
\noalign{\smallskip}
\multicolumn{1}{c}{Sample} & \multicolumn{1}{c}{$\Delta \log \left( \text{O}/\text{H} \right)$} & \multicolumn{1}{c}{$\Delta \log \left( \text{O}/\text{H} \right)$} & \multicolumn{1}{c}{$\Delta \log \left( \text{O}/\text{H} \right)$} \\    
 &  $\left( \left< \sigma_\mathrm{dist} \right>_\mathrm{VIPERS} \right)$ & $\left( \left< \sigma_\mathrm{med} \right>_\mathrm{VIPERS} \right)$ & (dex) \\
\hline                                   
\noalign{\smallskip}
    SDSS main sample & $0.198$ & $1.055$ & $0.0222$\\      
    SDSS p-control sample & $0.219$ & $1.078$ & $0.0241$\\
    SDSS d-control sample & $0.194$ & $1.040$ & $0.0249$ \\
    \noalign{\smallskip}
    \hline
\end{tabular*}
\end{table*}
\begin{table*}
\caption{Average (over bins) differences in metallicity between all SDSS-based and VIPERS not mass-complete samples (as in Fig.~\ref{fig:delta_fmr_masscomplete}) in $M_\star$-SFR bins. The differences are expressed in absolute units, in units of $\left < \sigma_\mathrm{dist} \right>_\mathrm{VIPERS}$, and in units of $\left < \sigma_\mathrm{med} \right>_\mathrm{VIPERS}$.}              
\label{tab:fmr_diff_notmasscomplete}      
\centering                                      
\begin{tabular*}{\textwidth}{l @{\extracolsep{\fill}} c c c}          
\hline    
\hline
\noalign{\smallskip}
\multicolumn{1}{c}{Sample} & \multicolumn{1}{c}{$\Delta \log \left( \text{O}/\text{H} \right)$} & \multicolumn{1}{c}{$\Delta \log \left( \text{O}/\text{H} \right)$} & \multicolumn{1}{c}{$\Delta \log \left( \text{O}/\text{H} \right)$} \\    
 &  $\left( \left< \sigma_\mathrm{dist} \right>_\mathrm{VIPERS} \right)$ & $\left( \left< \sigma_\mathrm{med} \right>_\mathrm{VIPERS} \right)$ & (dex) \\
\hline                                   
\noalign{\smallskip}
    SDSS main sample & $0.246$ & $2.205$ & $0.0314$\\      
    SDSS p-control sample & $0.250$ & $1.888$ & $0.0284$\\
    SDSS d-control sample & $0.262$ & $2.179$ & $0.0353$ \\
    \noalign{\smallskip}
    \hline
\end{tabular*}
\end{table*}

\begin{figure}
    \centering
    \resizebox{\hsize}{!}{\includegraphics{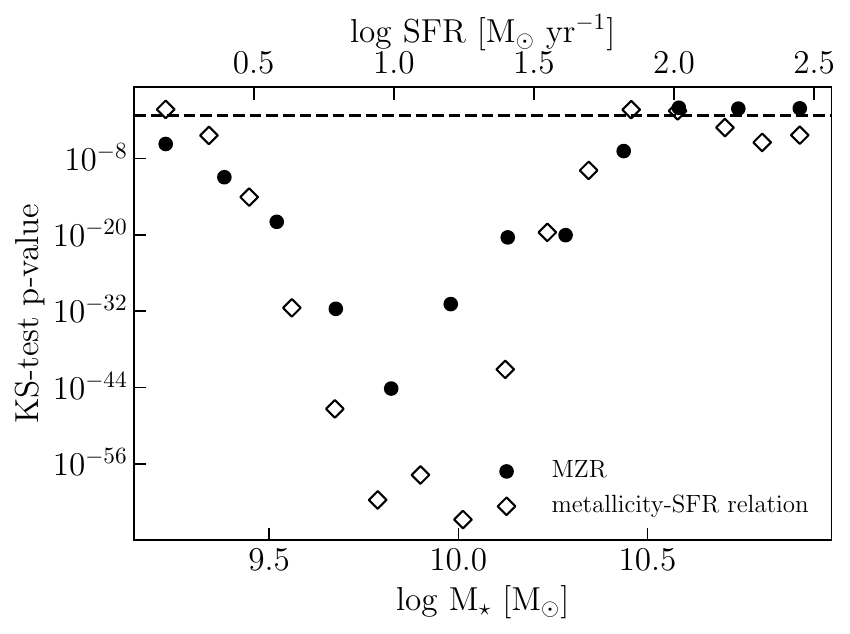}}
    \resizebox{\hsize}{!}{\includegraphics{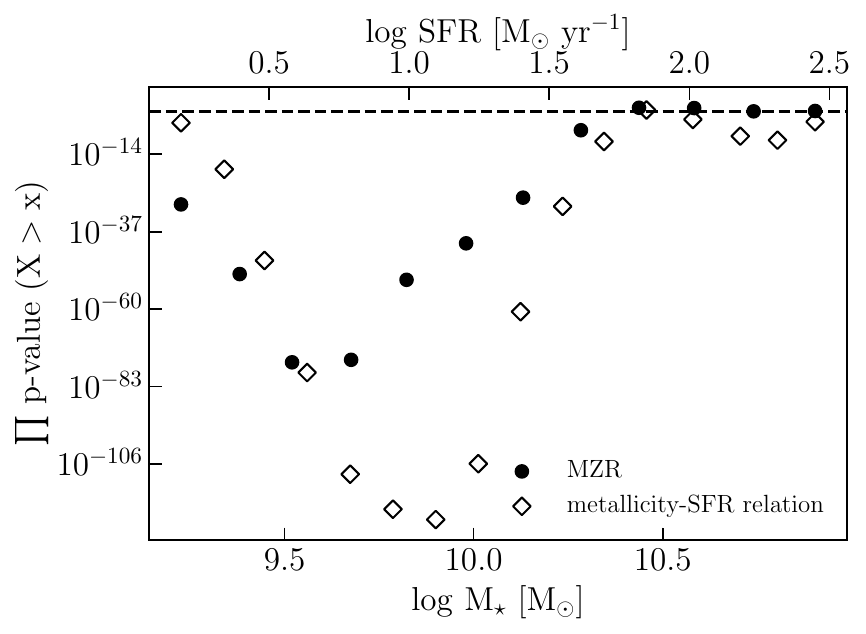}}
    \caption{Results of the KS-test between VIPERS and SDSS samples in order to check the evolution of the MZR and the metallicity-SFR relation. The top panel shows the p-value of the KS-test performed in $M_\star$ bins for the MZR (dots) and SFR bins for the metallicity-SFR relation (diamonds). The bottom panel is like the top panel but we report the product of the sequence for $\mathrm{X} > \mathrm{x}$ where X is the $M_\star$ or the SFR and x is the value on the x-axis. The dashed horizontal line ($y=0.05$) shows the threshold for the $95\%$ confidence level of the KS-test.}
    \label{fig:kstest}
\end{figure}

In order to quantify whether SDSS and VIPERS are statistically different, we perform a Kolmogorov-Smirnov test (KS-test).
We divide both samples in the same $M_\star$ or SFR bins and we perform the KS-test between the distributions of the samples within each bin.
Figure~\ref{fig:kstest} shows the resulting p-value as a function of the $M_\star$ and SFR.
Only the three highest $M_\star$, the fourth and fifth to last, and the lowest SFR bins have a p-value $\geq 0.05$, meaning the probability that the two samples are drawn by the same distribution is statistically significant ($\geq 95\%$).
Figure~\ref{fig:kstest} also shows the probability of having the samples drawn from the same distribution in $M_\star$ and SFR ranges.
Here, for the metallicity-SFR relation, only for ${\log \mathrm{SFR} \left[ M_\sun\ \mathrm{yr}^{-1} \right] \sim 1.75}$ the samples are statistically equivalent.
For the MZR, only for $\log M_\star \left[ M_\sun \right] \geq 10.4$ the samples are statistically equivalent.
From the point of view of the evolution of the MZR and metallicity-SFR relation, only the high $M_\star$ end of the MZR is comparable between low and intermediate redshifts, while the metallicity-SFR is statistically not comparable in the whole range explored.

\begin{figure}
    \centering
    \resizebox{\hsize}{!}{\includegraphics{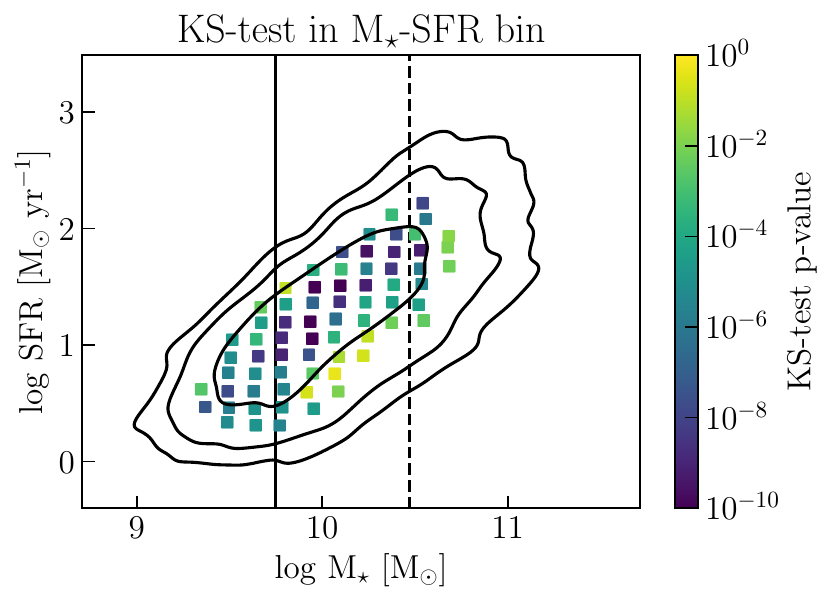}}
    \caption{Results of the KS-test between VIPERS and SDSS samples in order to check the evolution of the FMR. The scatter plot is color-coded according to the p-value resulting from the KS-test in $M_\star$-SFR bins. The color bar shows the two colors chosen above and below the threshold for the $95\%$ confidence level of the KS-test. The vertical black solid line indicates the mass below which the MZR at intermediate redshift flattens. The vertical black dashed line shows the most constrictive mass limit for completeness (${\log M_\star \left[ M_\sun \right] = 10.47}$ for ${0.65 < z <= 0.8}$) in the redshift range observed by VIPERS \citep{davidzon2016env}.}
    \label{fig:kstest_fmr}
\end{figure}

We also compare the metallicity distributions in $M_\star$-SFR bins in order to compare the FMR at low and intermediate redshifts by KS-test.
Figure~\ref{fig:kstest_fmr} shows the scatter in the $M_\star$-SFR plane color-coded according to the p-value resulting from the KS-test.
The majority of bins of the surface explored by the VIPERS sample show a p-value $\leq 10^{-4}$.
For the majority of the bins, the two samples are drawn from different distributions.
Even removing the area where the MZR at intermediate redshift flattens at low $M_\star$, the majority of bins still have really low values of the p-value.
Also, the evolution of the FMR is statistically significant between low and intermediate redshifts.
If we again apply a more strict data selection on $\mathrm{H}\beta$ ($\mathrm{S/N} > 5$), we can not accept the null hypothesis of the KS-test (data drawn from the same distribution) the number of bins for which the null hypothesis can be accepted is greatly reduced.

\begin{figure}
    \centering
    \resizebox{\hsize}{!}{\includegraphics{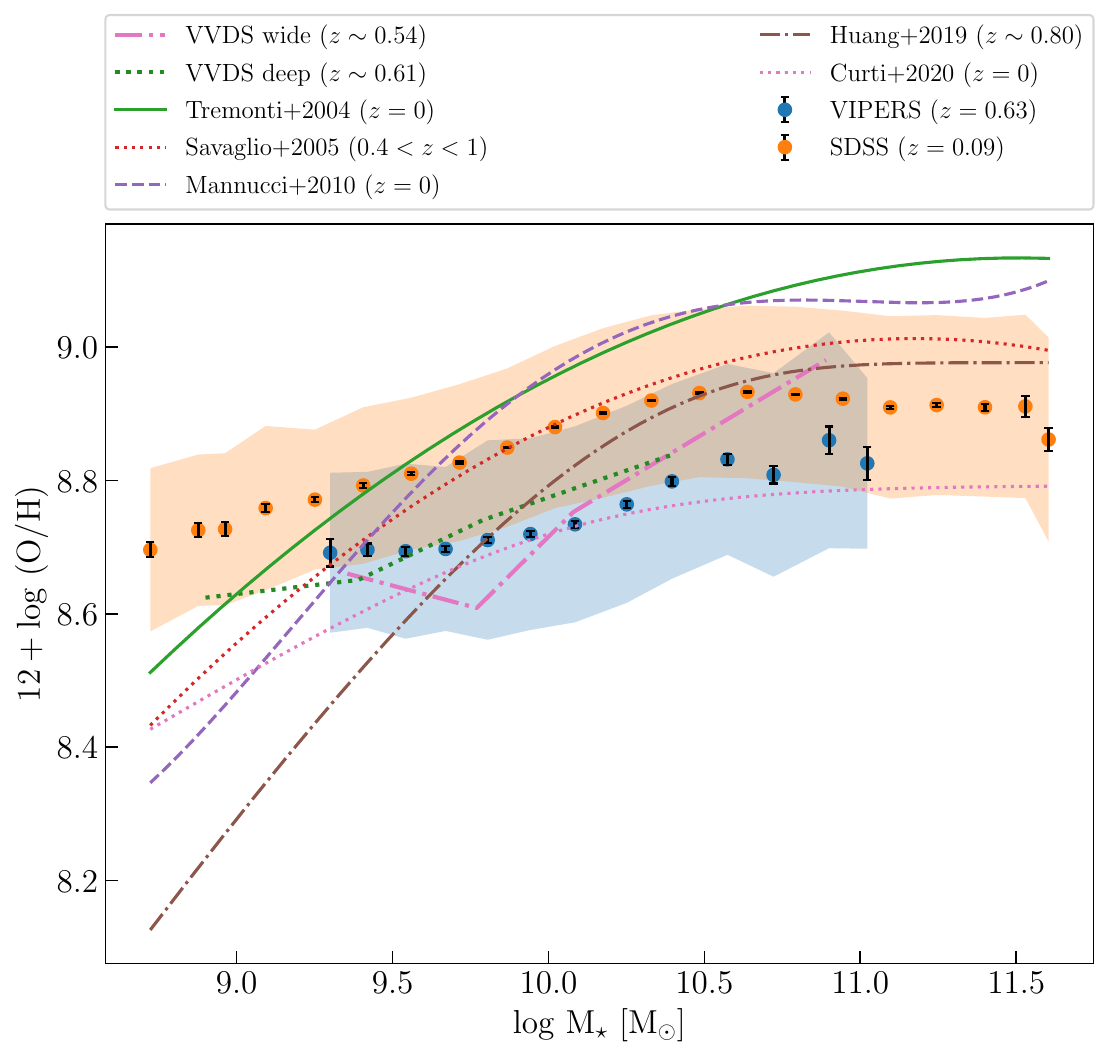}}
    \caption{Comparison of the MZR for the VIPERS (blue dots), the SDSS (orange dots), VVDS wide (dash-dotted pink line), and VVDS deep (dotted forest-green line) samples together with the fit functions in the literature. The shaded areas show the $1 \sigma_\mathrm{dist}$ while the black errorbars show the $1 \sigma_\mathrm{med}$ for the metallicity. The median redshifts are reported for the samples.}
    \label{fig:mzr_lit}
\end{figure}

Figure~\ref{fig:mzr_lit} shows the comparison between the MZR for the VIPERS and the SDSS samples with different fit reported in the literature \citep{tremonti2004origin, savaglio2005gemini, mannucci2010fundamental, huang2019mass, curti2020mass}.
We also report the fits of the MZR of studies claiming to measure the evolution of the MZR \citep{savaglio2005gemini, huang2019mass}.
The shape of the MZR for both samples agrees with the literature at high-$M_\star$ but the tail at low-$M_\star$ seems to be too much flatter in the VIPERS sample.

In the same plot, the MZR of the VVDS wide ($i_\mathrm{AB} < 22.5$) and deep ($i_\mathrm{AB} < 24.0$) fields are also reported.
These VVDS samples are cross-matched by the catalogs used by \cite{lamareille2009vimos} and \cite{lefevre2013vvds}.
Since we use here the VVDS sample only to validate the shape of the MZR, we do not recompute the physical properties using the values found in the catalog \citep{lamareille2009vimos}.
The MZR of the VIPERS sample follows the same shape as the VVDS samples showing a flattening at the low-$M_\star$ tail.
This behavior happens for $\log M_\star \left[ M_\sun \right] < 10.0$ where the VIPERS survey is not mass-complete while the completeness of the VVDS sample is ensured in the whole magnitude range of the spectroscopic survey \citep{mccracken2003virmos, lefevre2013vvds}.

\section{Discussion}\label{sec:disc}

In this paper, we studied the FMR at low redshift (SDSS data, median $z \sim 0.09$) and intermediate redshift (VIPERS data, median $z \sim 0.63$) using two different methods:
\begin{enumerate}[i)]
    \item the study of the FMR projections using also direct cross-matching between samples at different redshift;
    \item the non-parametric which compares the metallicity versus the normalized sSFR at different $M_\star$ bins cross-matching galaxies accordingly to the normalization of the sSFR.
\end{enumerate}
We aimed to study the influence of the methods on the conclusion about the evolution of the FMR.
At the same time,  we decided to check for the presence of observational biases that were not taken into account by cross-matching the catalogs at low and intermediate redshifts \citep{pistis2022bias}.
In this section, we discuss the results of the methods used for the comparison of the FMR between different redshift ranges (Sect.~\ref{sec:methods2}), the comparison of the samples (Sect.~\ref{sec:comparison2}), and the evolution of the MZR and the metallicity-SFR relation (Sect.~\ref{sec:evo2}).

\subsection{Methods of comparison}\label{sec:methods2}

In the study using the parametric method, we built two control samples with the following characteristics:
\begin{enumerate}[i)]
    \item cross-matching physical properties ($M_\star$ and SFR, p-control sample);
    \item reproducing the relative distance from the MS of the VIPERS sample (d-control sample).
\end{enumerate}

We used two different normalizations in the non-parametric method:
\begin{enumerate}[i)]
    \item with respect to the median sSFR of the sample at low redshift allowing us to compare galaxies with the same physical properties ($M_\star$ and SFR);
    \item with respect to the sSFR value from the fit of the MS allowing us to compare galaxies with the same relative distance from the MS at different redshift.
\end{enumerate}
This last method allows us to study the processes that move galaxies around the MS producing the intrinsic scatter of the MS itself via enhancement of the SFR or starvation of the galaxy.
We find that: 
\begin{enumerate}[i)]
    \item parametric method has the problem of inferring information about the FMR surface from the study of its median projections.
    In fact, these are expected to evolve compared to the whole surface. 
    Moreover, in other to compare samples with specific properties, it is necessary to cross-match the specific properties.
    \item non-parametric method has the advantage of being mostly independent of bias.
    Changing the normalization, this method allows us to compare galaxies with the same physical properties ($M_\star$ and SFR) or galaxies with the same relative distance from the MS between low and intermediate redshift.
\end{enumerate}



\paragraph{Advantages:} The non-parametric method is simpler to use than the various projections having the advantage of straightly comparing galaxies with similar physical properties or relative distance from the MS without the necessity to cross-match the catalogs.
They are also independent of the biases that could be introduced by observation and data selection.
The non-parametric method, using the normalization from the MS, gives also information about the processes that lead to a drop or an enhancement of the SFR.

\paragraph{Disadvantages:} The parametric method needs to be taken with high caution as it can be affected by biases introduced by observation and data selection as the FMR projections are sensitive to these kinds of biases, (combining the $M_\star$ and SFR reduces the effects due to biases).
Since the FMR projections are expected to evolve compared to the FMR itself, it can be difficult to deduce information on the whole FMR starting from its projections. 
To compare specific properties between different samples, it is necessary to cross-match these properties.

\subsection{Comparison of FMR between different redshifts ranges}\label{sec:comparison2}

Being particularly careful to homogenize the samples at different redshifts, the expected evolution of the MZR (Fig.~\ref{fig:fmr_proj}, upper left panel) is statistically significant compared with the $\sigma_\mathrm{med}$ on metallicity.
This evolution is confirmed by the KS-test (Sect.~\ref{sect:mzrevo}, Fig.~\ref{fig:kstest}).

The metallicity difference between SDSS-based and VIPERS samples (Table~\ref{tab:fmr_diff}) in $M_\star$-SFR bins does not show any particular variation for different SDSS-based samples but shows a systematic with $M_\star$ which is dominated by the flattening at low $M_\star$ present in the sample at intermediate redshift.

The study of the slopes in the normalized (by the value expected by the MS) metallicity-sSFR relation (Fig.~\ref{fig:comp_fmr_relms_ssfr}) shows an increasing difference between samples at intermediate and low redshift with $M_\star$.
The same shift in $M_\star$ was also found by \cite{salim2015mass} at $z \sim 2.3$ based on \cite{steidel2014kbss}  from the Keck Baryonic Structure Survey (KBSS). 

The study of the slopes in the normalized (by the value expected from the MS) metallicity-sSFR relation (Fig.~\ref{fig:comp_fmr_relms_ssfr}) allows us to study the processes that move the galaxies around the MS.
Galaxies above the MS (${\delta \log \text{sSFR} > 0}$, Fig.~\ref{fig:slope} upper panel) undergo processes enhancing the SFR.
From the point of view of gas-inflow, the small dilution (shallower slope) at intermediate redshift suggests an advanced stage of evolution of the infalling gas, with metallicity closer to the ISM.
The small metallicity difference between ISM and infalling gas can be explained by assuming a less processed ISM or a more processed infalling gas.
The first scenario does not seem to be accurate since the metallicity of galaxies at intermediate redshift is not (statistically significant) different from those at median $z \sim 0.09$.
According to the hierarchical model of galaxy formation, the merging rate increases with redshift assuming closer galaxies inside the clusters.
This already ``metal-rich'' gas can be previously expelled into the intracluster medium (ICM), suggesting environmental effects.
This also suggests that the assumption of the pristine nature of the infalling gas is not
always true.

The difference in metal dilution between the VIPERS and SDSS samples is more prominent at low-$M_\star$, but such an effect is less prominent towards higher $M_\star$.
It has been shown that metallicity dilution can be increasingly significant in the case of gaseous mergers \citep[both major and minor, e.g.,][]{ellison2013merger}.
\cite{bustamante2020mergers} show that mergers cause a large scatter of FMR and the trends continue to the post-merger stage.
Similar conclusions were obtained by theoretical studies that utilize idealized hydro simulations \citep[e.g.,][]{bustamante2018sim}.
Although a detailed analysis of galaxy merger impact on estimated FMR is out of the scope of the present paper, the fact that galaxy merging rate rises with redshift \citep[e.g.,][]{ventou2017merging} may partly explain the difference between the slopes inferred from VIPERS and SDSS data.

Below the main sequence (${\delta \log \text{sSFR} < 0}$, Fig.~\ref{fig:slope} bottom panel), the intermediate redshift sample shows, compared to the low redshift sample, ongoing processes that are diluting more significantly the metals in the whole $M_\star$ range.
Again the difference at different redshifts decreases with increasing $M_\star$.
In this case, the situation is inverted at low-$M_\star$ with the low redshift sample having weaker dilution of the metals.

On the one hand, it can reflect the dilution-starvation scenario proposed in recent studies \citep[e.g., more recent star formation dilutes metals more efficiently, while suppression of fresh gas leads to enhanced metallicity and lower SFR,][]{kumari2021fmr}.
On the other hand, among the processes that can happen to quench galaxy SFR (${\delta \log \text{sSFR} < 0}$) there are the so-called dry merger \citep{bell2006merger, khochfar2009merger}.
These dry merging events are characterized by a low amount of gas, suggesting the participation of older galaxies already in an advanced stage of evolution, and they occur for massive galaxies ($\log M_\star \left[ M_\odot \right] \geq 10.4$).
The reduction in dilution, corresponding to a shallower slope, can be explained by dry merging events.
From the point of view of outflows instead, a more negative slope at small $M_\star$ suggests a higher efficiency in removing the metals from the ISM with the main production of metals in the bulge of the galaxy or a bigger amount of gas expelled from the galaxy itself (starvation).

Our study does not aim at quantifying dust masses of the different samples used, mainly due to the lack of infrared detections for the majority of these galaxies.
However, dust content in galaxies plays a major role in the evolution of the ISM, and a driver of the SFR.
Higher redshift galaxies tend to have larger dust reservoirs \citep[e.g.,][]{Takeuchi2005dust, Whitaker2017dust}.
Additionally, galaxies with higher $M_\star$ have larger dust masses \citep{Beeston2018dust}.
Metals can be converted into dust during the complex evolution of the ISM.
Therefore, Fig.~\ref{fig:slope} can be seen as a metal depletion process by the current content of dust.
In fact, for actively SF galaxies (${\delta \log \text{sSFR} > 0}$), the similar depletion efficiency of metals at higher $M_\star$ at different redshifts, can potentially be explained by the supposed larger dust content towards the higher $M_\star$.
For lower $M_\star$, metal depletion of VIPERS galaxies is weaker than that of SDSS.

For the less SF galaxies (${\delta \log \text{sSFR} < 0}$), the metal depletion efficiency at low redshift is weaker than at higher redshift for the whole range of $M_\star$.
The difference is likely to be driven by a more efficient dust-to-metal ratio and a higher fraction of available cold gas \citep{devis2019metaldustpedia}.
Indeed, very recent studies of dust-to-metal co-evolution in galaxies at intermediate redshifts ($z < 0.7$) found that the conversion of metals to dust can be efficient even in evolved systems with old stellar ages \citep{donevski2023dust}.
Our finding displayed in Fig.~\ref{fig:slope} qualitatively agrees with this scenario.

All the aforementioned physical aspects of galaxies are responsible for metal depletion at different redshift ranges.
Even though dark matter and galaxy environment are responsible for shaping the evolution of galaxies, dust is crucial in the interaction between the ISM components.
The metallicity plays an important role in dust content and consequently dust attenuation \citep{shivaei2020irxbeta, casasola2022dustpedia, pantoni2021dust, Hamed2023attenuation}, along with the dependence on the environment \citep{hamed2023irxbeta}.

The reduced change between different redshifts at higher $M_\star$ can be explained by the dark matter halo bias.
This bias leads to a faster evolution of massive galaxies.
For the same reason, the MZR and FMR flatten at high $M_\star$.

\subsection{Evolution of the MZR and metallicity-SFR relation}\label{sec:evo2}

Looking for the most comparable and homogeneous measurements of the galaxy properties, we observe a small shift between the low redshift and intermediate redshift samples.
This evolution is not statistically significant with respect to the $\sigma_\mathrm{med}$ on metallicity ($\Delta \log \left( \text{O}/\text{H} \right) < 2 \left< \sigma_\mathrm{med} \right>_\mathrm{VIPERS}$ in $M_\star$-SFR bins).
However, we do not observe any evolution of the scatter around the relation with respect to the $\sigma_\mathrm{dist}$ from intermediate and low redshift.

Once the intermediate sample is divided into two redshift bins, ${0.48 < z < 0.65}$ and ${0.65 < z < 0.80}$, the MZR within the redshift bins overlaps entirely with the MZR using the whole sample (Fig.~\ref{fig:evo} upper panel).
Once taken into account the stellar mass-completeness, the VIPERS mass-complete sub-sample follows the same MZR as the main VIPERS sample.
Instead, the metallicity-SFR relation is slightly more sensitive to both redshift and mass-completeness.
The differences remain within the $\sigma_\mathrm{dist}$ in metallicity but not within the $\sigma_\mathrm{med}$.

From the point of view of the FMR, the metallicity difference is reduced by about half.
Then, stellar mass-completeness is not an important property of the catalog in the MZR, but it is an important property in the metallicity-SFR relation or FMR.
The stellar mass-completeness affects the metallicity-SFR relation by changing the correlation coefficient by $\sim -0.3$ and reducing the difference between different SFR calibrations (see Appendix~\ref{app:sfr}).
The stellar mass-completeness condition reduces the mass range explored by the VIPERS sample around the position of the peak of the SDSS sample.
The SFR and metallicity distributions are shifted towards higher values once the condition is applied.
This is reflected in the FMR with a smaller difference between SDSS and VIPERS samples (see Sect.~\ref{sect:mzrevo}).
Previous studies \citep{mannucci2010fundamental, cresci2012zcosmos, maier2015fmr, salim2015mass, gao2018mass, sanders2021mosdefevo, strom2022kbss, curti2023jades} did not take into account the stellar mass-completeness of the catalogs.
The evolution of the FMR can be scaled down once this property of the catalogs is considered.
Also, the KS-tests performed on the distribution of the samples at low and intermediate redshift within $M_\star$, SFR, and $M_\star$-SFR bins confirm the evolution of the MZR and FMR (Sect.~\ref{sect:mzrevo}, Fig.~\ref{fig:kstest}, and Fig.~\ref{fig:kstest_fmr}).

\section{Conclusions}\label{sec:concl}

We check if the so-called unified or fundamental relations on metallicity stand at different redshifts, under various selection criteria, and methods of comparison.
Our analysis focused on the FMR and its behavior at low (SDSS data, median $z \sim 0.09$) and intermediate redshifts (VIPERS data, median $z \sim 0.63$).
We found a systematic shift in metallicity at median redshift $z \sim 0.63$ compared to the local Universe.
The average metallicity difference in $M_\star$-SFR bins is ${\Delta \log \left( \mathrm{O/H}\right) \sim 1.8 \left< \sigma_\mathrm{med} \right>_\mathrm{VIPERS}}$, which is statistically significant according to the performed KS-test.
Despite the small difference can be dominated by systematic uncertainties in the calculation of SFR and metallicities (e.g., the $f_{factor}$, the attenuation parameter, the extinction law, the used SFH in SED fitting), we minimized the systematics by computing the physical properties most homogeneously.
This study points out a hint of an earlier evolution of the FMR than expected.
Future surveys, e.g. Euclid space mission, will be able to confirm this evolution once the uncertainties can be reduced.

In mass-complete samples, the metallicity difference is reduced to ${\Delta \log \left( \mathrm{O/H}\right) \sim 1 \left< \sigma_\mathrm{med} \right>_\mathrm{VIPERS}}$.
This conclusion may be influenced by the limited parameter space spanned, being the lower stellar mass galaxies where the difference is larger out from the analysis.
A careful reading of the results, and their underlying selection criteria, are crucial in studies of the mass-metallicity and fundamental metallicity relations.
Another effect of having a mass-complete sample is the change in the value of the correlation coefficient between metallicity and SFR when some calibration (e.g., based on $\mathrm{H}\beta$ or $\left[ \ion{O}{ii} \right]$ once the metallicity correction is applied) are used.

The reduction of metal dilution of galaxies below the MS (${\delta \log \text{sSFR} < 0}$) at high $M_\star$ suggests processes of dry mergers at work, instead above the MS (${\delta \log \text{sSFR} > 0}$) an inflow of metal-rich gas with metallicity close to the ISM is suggested at intermediate redshift.
On average, the difference in metal dilution is decreasing with stellar mass at larger redshifts.

When studying the mass-metallicity and fundamental metallicity relations, we recommend using the non-parametric approach which provides similar results compared to parametric prescriptions, being easier to use and results fair to interpret. The non-parametric methodology provides a convenient way to compare physical properties, with a smaller impact on observational selection biases.


\begin{acknowledgements}

This research was supported by the Polish National Science Centre grant UMO-2018/30/M/ST9/00757. K.M. is grateful for support from the Polish National Science Centre via grant UMO-2018/30/E/ST9/00082. M.F. acknowledges support from the First TEAM grant of the Foundation for Polish Science No. POIR.04.04.00-00-5D21/18-00 (PI: A. Karska). D.D. acknowledges support from the National Science Centre (grant SONATA-16, UMO-2020/39/D/ST9/00720). M.H. acknowledges the support of the National Science Centre (UMO-2022/45/N/ST9/01336). W.J.P. has been supported by the Polish National Science Center project UMO-2020/37/B/ST9/00466 and by the Foundation for Polish Science (FNP). M.R. acknowledges support from the Narodowe Centrum Nauki (UMO-2020/38/E/ST9/00077) and support from the Foundation for Polish Science (START 063.2023). This paper uses data from the VIMOS Public Extragalactic Redshift Survey (VIPERS). VIPERS has been performed using the ESO Very Large Telescope, under the ``Large Programme'' 182.A-0886. The participating institutions and funding agencies are listed at \url{http://vipers.inaf.it}. This research uses data from the VIMOS VLT Deep Survey, obtained from the VVDS database operated by Cesam, Laboratoire d'Astrophysique de Marseille, France. This research made use of Astropy\footnote{\url{http://www.astropy.org}}, a community-developed core Python package for Astronomy. 
\end{acknowledgements}

\bibliographystyle{aa}
\bibliography{Bibliography}

\begin{appendix}

\section{Validation of line measurements}\label{app:val}

\begin{figure}[h!]
    \centering
    \resizebox{\hsize}{!}{\includegraphics{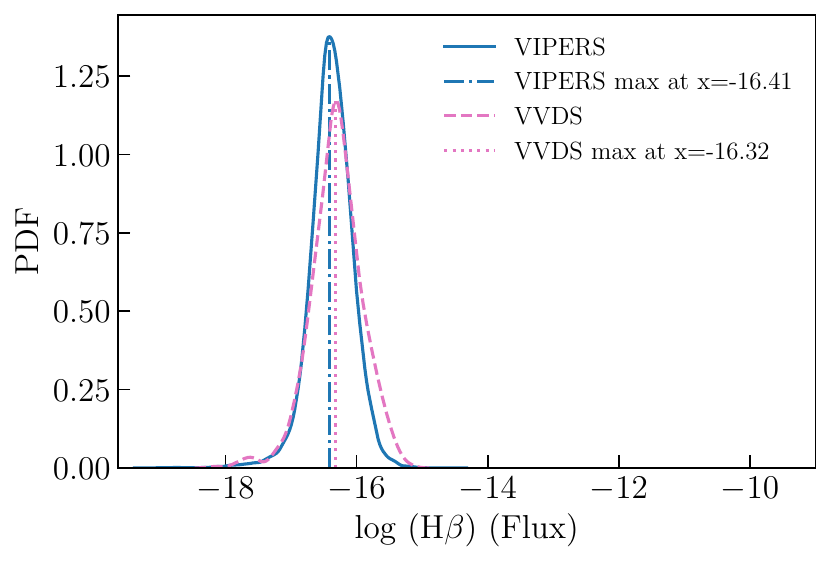}\includegraphics{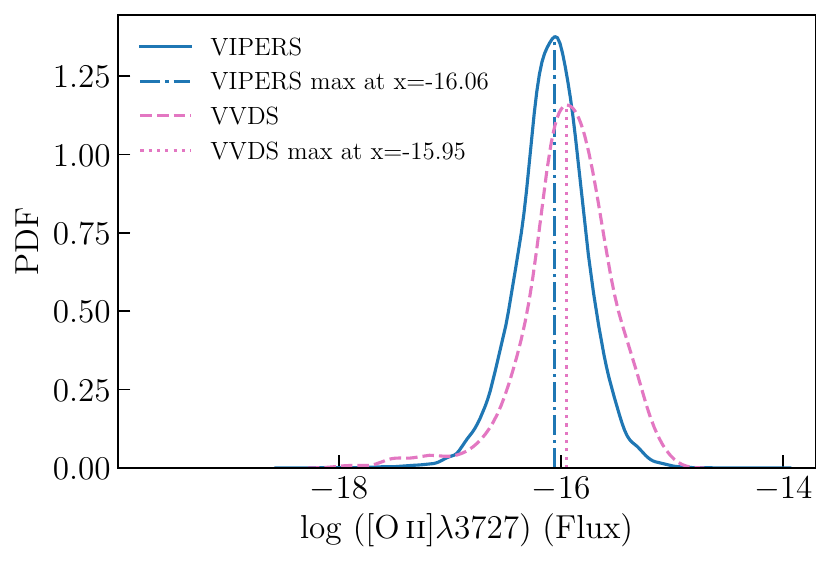}}
    \resizebox{\hsize}{!}{\includegraphics{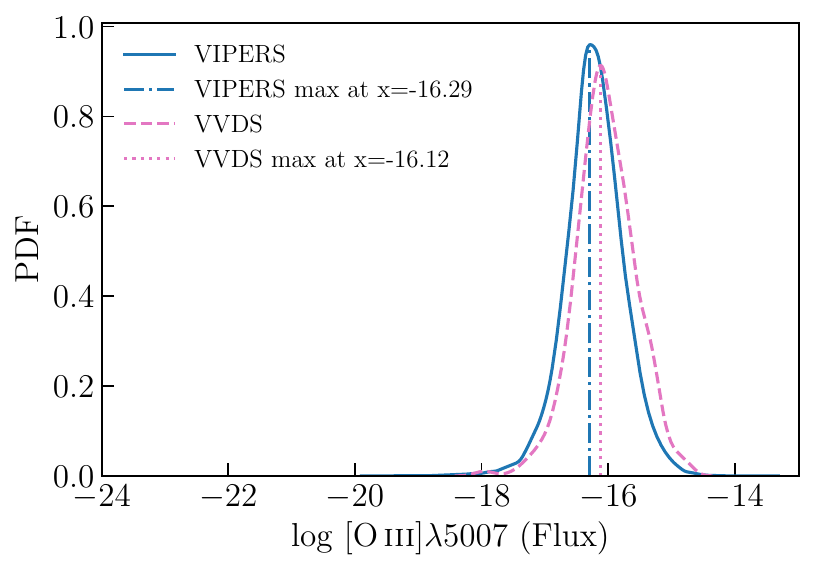}\includegraphics{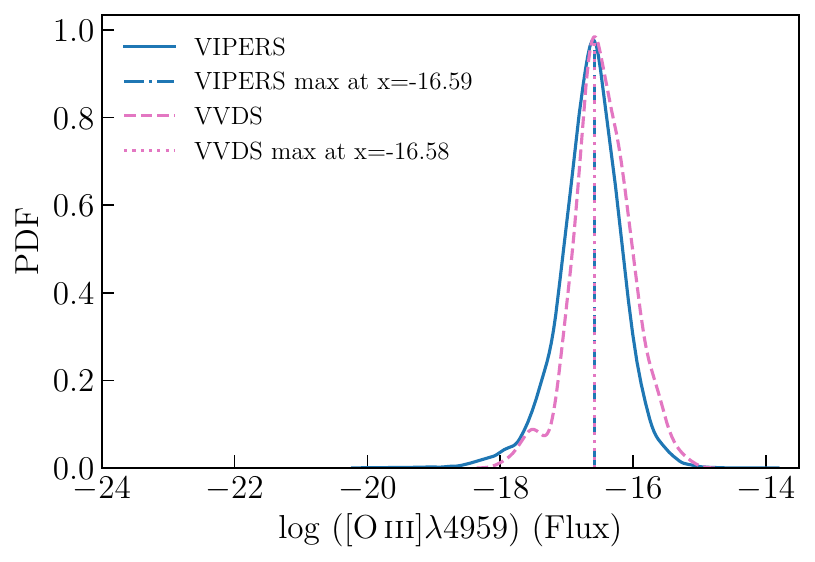}}
    \caption{Comparison of flux distributions for $\text{H}\beta$ (upper left), $\left[ \ion{O}{ii} \right]\lambda 3727$ (upper right), $\left[ \ion{O}{iii} \right]\lambda 5007$ (bottom left), and $\left[ \ion{O}{iii} \right]\lambda 4959$ (bottom right) lines between VIPERS (blue solid line) and VVDS (pink dashed line) samples. In the same plot is highlighted the position of the maximum of each distribution.}
    \label{fig:vvds_fluxcomp}
\end{figure}

\begin{figure}[h!]
    \centering
    \resizebox{\hsize}{!}{\includegraphics{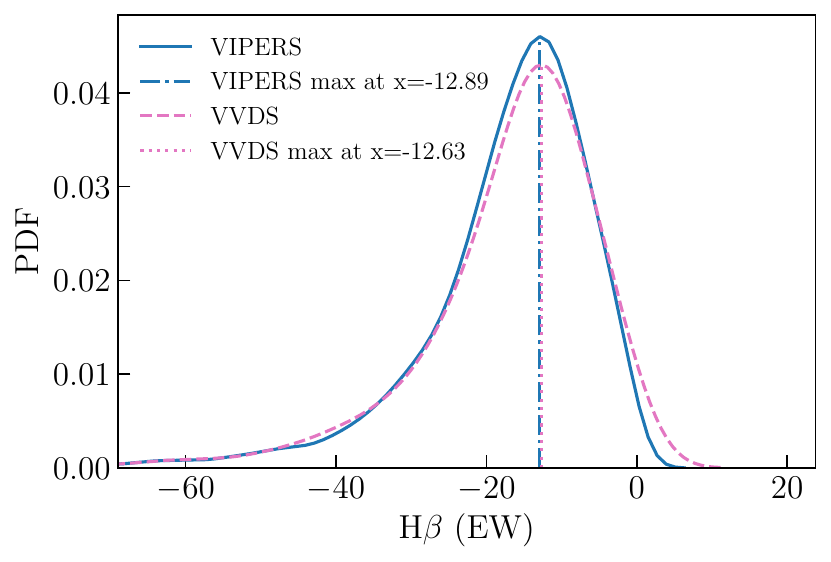}\includegraphics{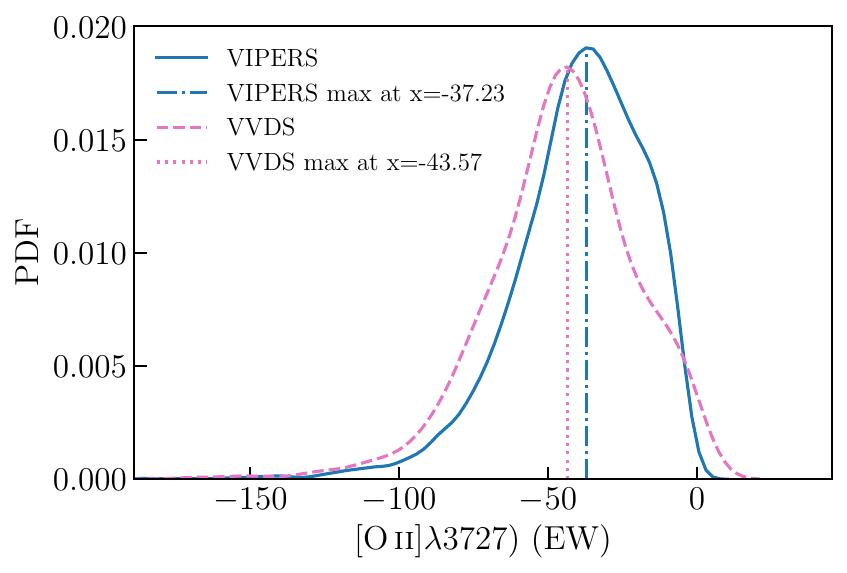}}
    \resizebox{\hsize}{!}{\includegraphics{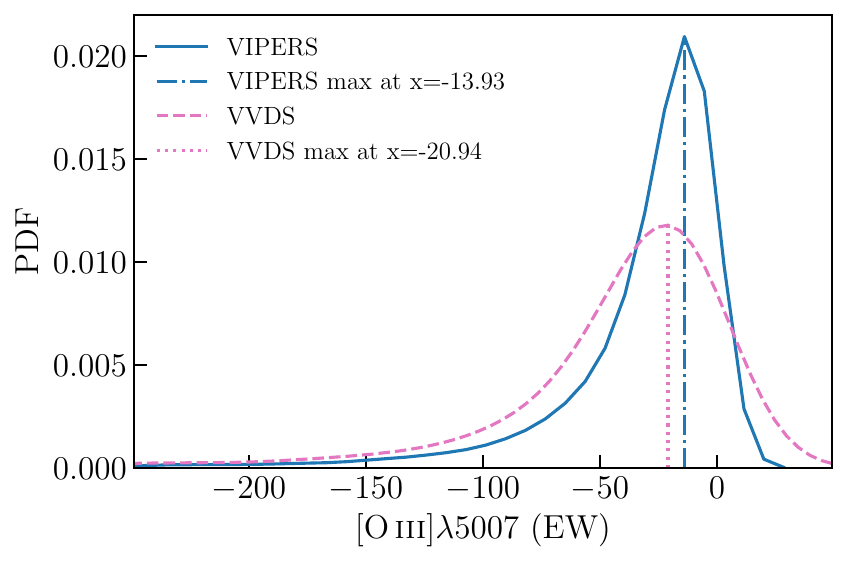}\includegraphics{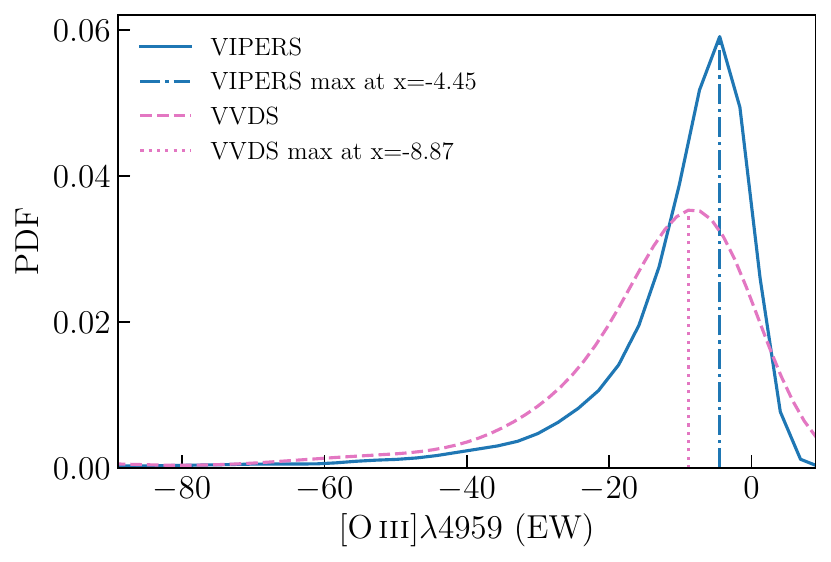}}
    \caption{Comparison of EW distributions for $\text{H}\beta$ (upper left), $\left[ \ion{O}{ii} \right]\lambda 3727$ (upper right), $\left[ \ion{O}{iii} \right]\lambda 5007$ (bottom left), and $\left[ \ion{O}{iii} \right]\lambda 4959$ (bottom right) lines between VIPERS (blue solid line) and VVDS (pink dashed line) samples. In the same plot, the position of the maximum of each distribution is highlighted.}
    \label{fig:vvds_ewcomp}
\end{figure}

We compare here directly the distributions of the fluxes and EWs for all the lines used in this paper between VIPERS and VVDS samples.
Figure~\ref{fig:vvds_fluxcomp} and Fig.~\ref{fig:vvds_ewcomp} show the comparison of the distributions of fluxes and EWs, respectively, for both VIPERS and VVDS samples.

\FloatBarrier

Table~\ref{tab:fluxcomparison} summarizes the comparison between the distribution of the fluxes.
Table~\ref{tab:ewcomparison} summarizes the comparison between the distribution of the EWs.
We find a good agreement between the flux and EW distributions for all the lines in exam according to their median, MAD, and position of the peaks.
We also performed a KS-test for the null hypothesis that two samples were drawn from the same distribution. We choose a confidence level of $95\%$; that is, we will reject the null hypothesis in favor of the alternative if the p-value is less than $0.05$ which we report in Table~\ref{tab:kscomparison}.
\begin{table}
\caption{Statistic of the comparison between the distributions.}              
\label{tab:fluxcomparison}      
\centering                                      
\begin{tabular*}{\columnwidth}{l @{\extracolsep{\fill}} c c c c}          
\hline    
\hline
\noalign{\smallskip}
 Emission line & \multicolumn{2}{c}{Median flux} & \multicolumn{2}{c}{MAD flux}\\    
 & VIPERS & VVDS & VIPERS & VVDS\\
 \hline
\noalign{\smallskip}
    $\text{H}\beta$ & $5.56$ & $4.95$ & $1.78$ & $2.44$\\      
    $\left[ \ion{O}{ii} \right]\lambda 3727$ & $8.35$ & $12.04$ & $3.57$ & $5.68$\\
    $\left[ \ion{O}{iii} \right]\lambda 5007$ & $6.04$ & $8.52$ & $3.52$ & $5.13$\\
    $\left[ \ion{O}{iii} \right]\lambda 4959$ & $2.39$ & $3.21$ & $1.42$ & $1.80$\\
    \noalign{\smallskip}
    \hline
\end{tabular*}
\tablefoot{The fluxes and MADs are expressed in units of $10^{-17}\text{erg}\text{ s}^{-1}\text{cm}^{-2}$.}
\end{table}
\FloatBarrier
\begin{table}[h!]
\caption{Statistic of the comparison between the distributions.}              
\label{tab:ewcomparison}      
\centering                                      
\begin{tabular*}{\columnwidth}{l @{\extracolsep{\fill}} c c c c }          
\hline    
\hline
\noalign{\smallskip}
 Emission line & \multicolumn{2}{c}{Median EW} & \multicolumn{2}{c}{MAD EW}\\    
 & VIPERS & VVDS & VIPERS & VVDS\\
 \hline
\noalign{\smallskip}
    $\text{H}\beta$ & $-14.97$ & $-14.33$ & $5.98$ & $5.80$\\      
    $\left[ \ion{O}{ii} \right]\lambda 3727$ & $-37.29$ & $-44.66$ & $14.13$ & $14.38$\\
    $\left[ \ion{O}{iii} \right]\lambda 5007$ & $-19.45$ & $-26.85$ & $12.41$ & $17.11$\\
    $\left[ \ion{O}{iii} \right]\lambda 4959$ & $-7.03$ & $-10.57$ & $4.57$ & $6.04$ \\
    \noalign{\smallskip}
    \hline
\end{tabular*}
\tablefoot{The EWs and MADs are expressed in units of {\AA}.}
\end{table}
\FloatBarrier
\begin{table}[h!]
\caption{Statistic of the comparison between the distributions.}              
\label{tab:kscomparison}      
\centering                                      
\begin{tabular*}{\columnwidth}{l @{\extracolsep{\fill}} c c}          
\hline    
\hline
\noalign{\smallskip}
 Emission line & \multicolumn{2}{c}{KS-test p-value} \\    
 & Flux & EW\\
 \hline
\noalign{\smallskip}
    $\text{H}\beta$ & $8.57\times 10^{-8}$ & $0.30$\\      
    $\left[ \ion{O}{ii} \right]\lambda 3727$ & $9.71\times 10^{-31}$ & $1.57 \times 10^{-13}$\\
    $\left[ \ion{O}{iii} \right]\lambda 5007$ & $7.55 \times 10^{-17}$ & $5.16 \times 10^{-12}$\\
    $\left[ \ion{O}{iii} \right]\lambda 4959$ & $1.65\times 10^{-17}$ & $6.12\times 10^{-17}$\\
    \noalign{\smallskip}
    \hline
\end{tabular*}
\end{table}
\FloatBarrier
According to the p-value of the KS-test only the distributions for $\mathrm{H}\beta$ EW is consistent with the null hypothesis, in the other cases the null hypothesis must be rejected.

\section{Signal to noise of line measurements}\label{app:hists}

\begin{figure}[h!]
    \centering
    \resizebox{\hsize}{!}{\includegraphics{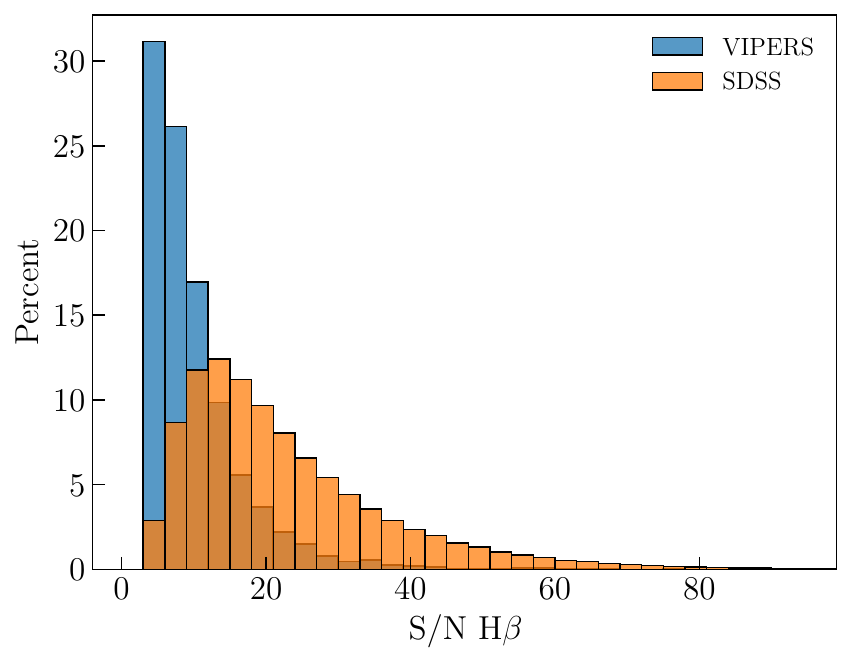}\includegraphics{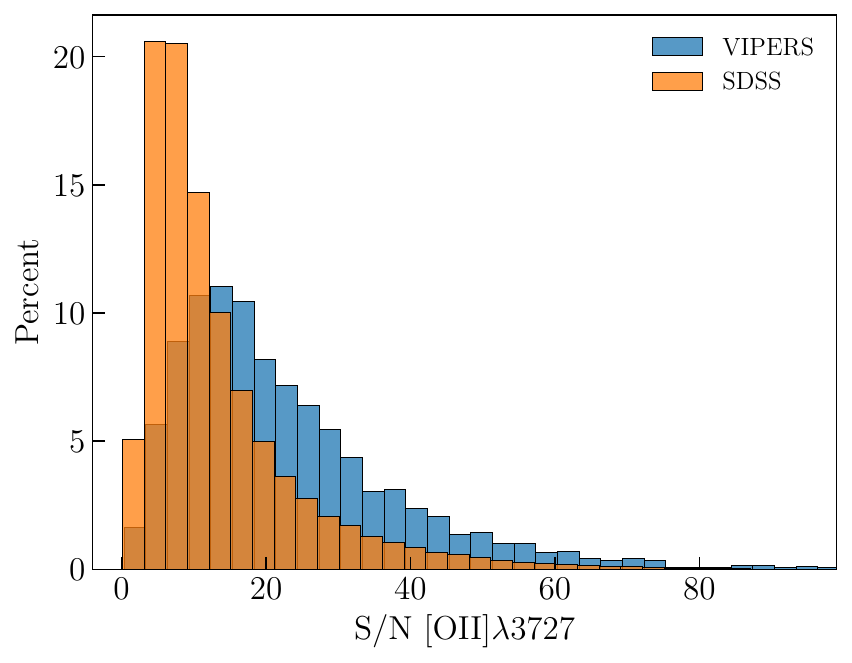}}
    \resizebox{\hsize}{!}{\includegraphics{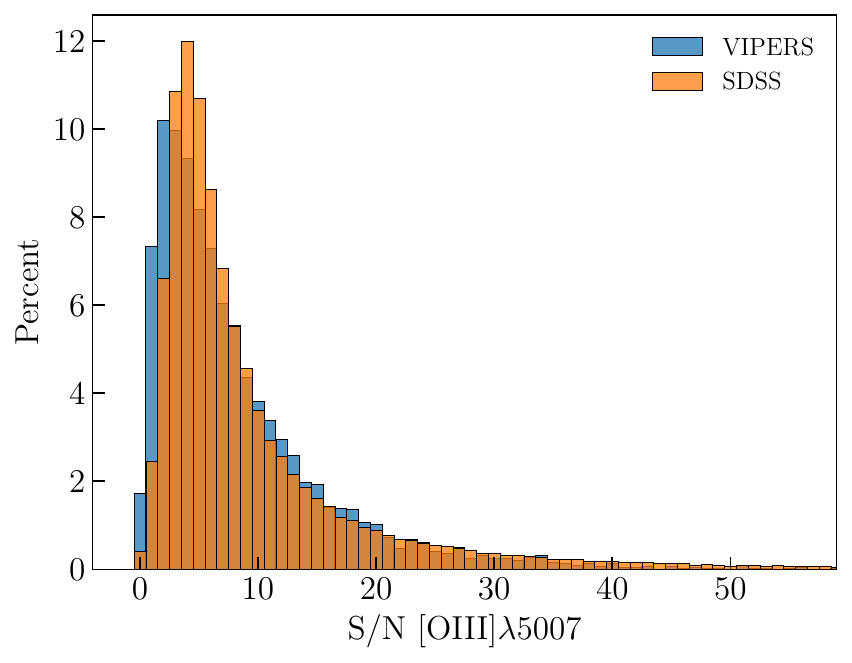}\includegraphics{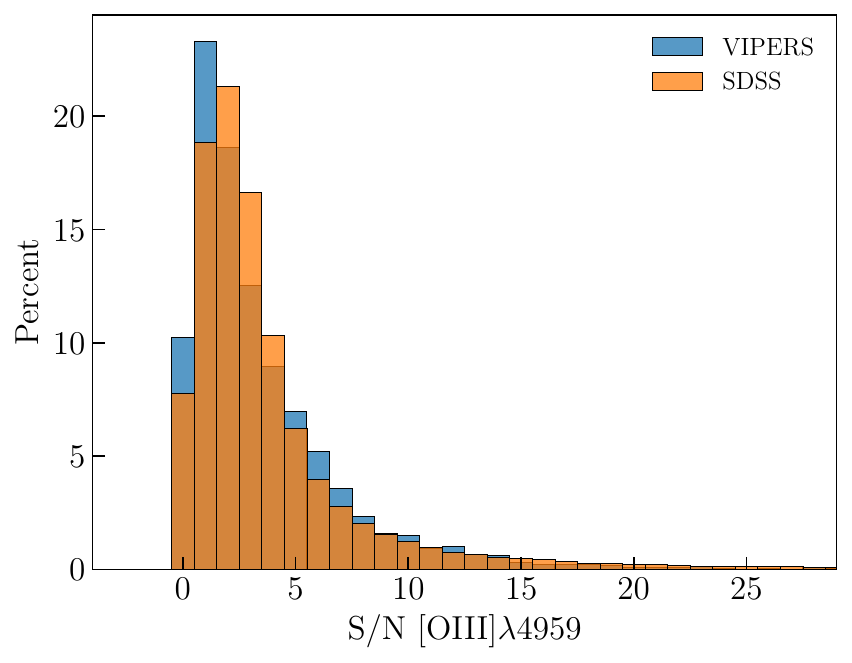}}
    \caption{Histograms of the S/N of the flux of each line normalized so that bar heights sum up to 100 for VIPERS (blue) and SDSS (orange) samples.}
    \label{fig:hist}
\end{figure}

Figure~\ref{fig:hist} shows the histograms of the S/N of the flux of each line, normalized so that bar heights sum up to 100.
The emission lines used in this study have similar distributions as SDSS, where the selection on H$\alpha$ is applied as in \cite{curti2020mass}.
\FloatBarrier

\section{Estimation of the attenuation}\label{app:av}

\begin{figure}[h!]
    \centering
    \resizebox{\hsize}{!}{\includegraphics{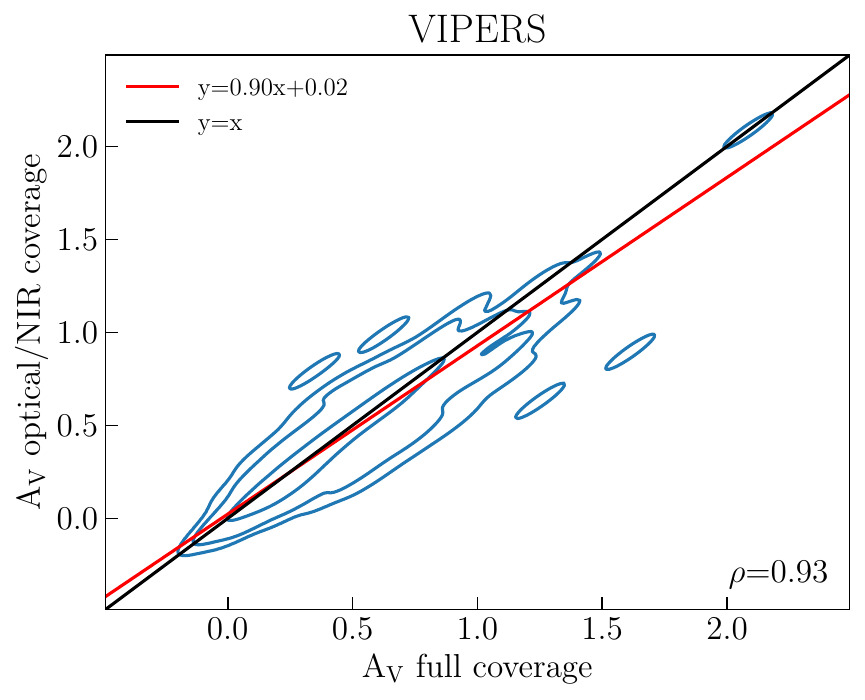}}
    \caption{$A_\mathrm{V}$ estimation from optical/NIR versus the $A_\mathrm{V}$ up to HERSCHEL/SPIRE band.}
    \label{fig:av}
\end{figure}

Because of the limited wavelength coverage for the VIPERS sample (up to optical/NIR), the estimation of the attenuation can be problematic.
We then compare the attenuation computed via SED fitting for a subsample of $\sim 600$ SF galaxies for which the coverage goes up to the HERSCHEL/SPIRE band.
Figure~\ref{fig:av} shows the estimations using only the optical/NIR bands versus the estimations going up with the wavelength coverage to HERSCHEL/SPIRE band.
We found a good agreement between the attenuation computed up to the optical/NIR infrared and the values computed up to HERSCHEL/SPIRE.
\FloatBarrier

\section{Effects of different SFR calibrations}\label{app:sfr}

The SFR can be estimated via different calibrations using different emission lines or applying a metallicity correction.
For example, the $\left[ \ion{O}{ii} \right]$ flux also depends on metallicity.
A more metal-independent calibration uses the $\mathrm{H}\beta$ flux to estimate the $\mathrm{H}\alpha$ and then use the \cite{kennicutt1998star} to calculate the SFR.

Table~\ref{tab:sfrcorr} shows the correlation coefficients between SFR calibrations and the metallicity.
An anti-correlation between SFR and metallicity is expected from both observations \citep{lara2010plane, mannucci2010fundamental, mannucci2011metallicity, yates2012relation, andrews2013mass, bothwell2013fmr, salim2014critical, zahid2014mzr, yabe2015fmr, hunt2016plane, sanders2018mosdef, cresci2019fundamental, curti2020mass, sanders2021mosdefevo} and theoretical models \citep{ellison2007clues, dave2011galaxy, yates2012relation, lilly2013fmr, peng2014evo, delucia2020mzr}.
For the VIPERS sample, only the calibration using the $\left[ \ion{O}{ii} \right]$ flux, without metallicity correction, shows an anti-correlation.
The situation results better if we consider the VIPERS mass-complete sub-sample having a much closer correlation coefficient with the SDSS sample.
\begin{table}[h!]
\caption{Correlation coefficients between SFR calibrations and metallicity.}              
\label{tab:sfrcorr}      
\centering                                      
\begin{tabular*}{\columnwidth}{l @{\extracolsep{\fill}} c c c c}          
\hline    
\hline
\noalign{\smallskip}
 SFR & VIPERS & VIPERS & VIPERS & SDSS\\    
 calibration & & MC & not MC & \\
\noalign{\smallskip}
 \hline
\noalign{\smallskip}
    $\left[ \ion{O}{ii} \right]$ & $-0.15$ & $-0.49$ & $-0.21$ & $-0.40$ \\ 
    $\left[ \ion{O}{ii} \right]_\mathrm{c}$ & $+0.17$ & $-0.09$ & $+0.10$ & $-0.13$ \\ 
    $\mathrm{H}\beta$ & $+0.16$ & $-0.13$ & $+0.10$ & $-0.19$ \\
    \noalign{\smallskip}
    \hline
\end{tabular*}
\tablefoot{MC stands for ``mass complete''. $\left[ \ion{O}{ii} \right]_\mathrm{c}$ stands for the SFR calculated from the line $\left[ \ion{O}{ii} \right]$ with the metallicity correction applied.}
\end{table}
\FloatBarrier

\begin{figure}[h!]
    \centering
    \resizebox{\hsize}{!}{\includegraphics{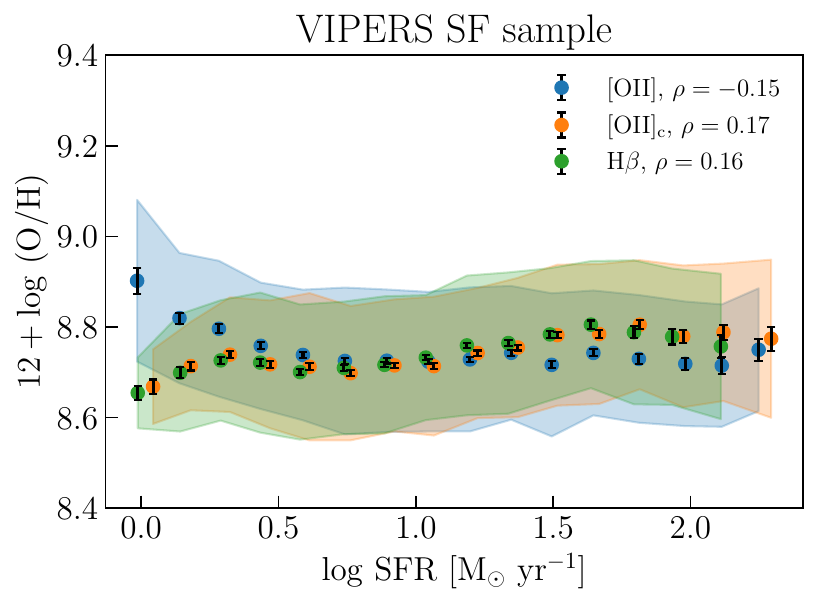}}
    \resizebox{\hsize}{!}{\includegraphics{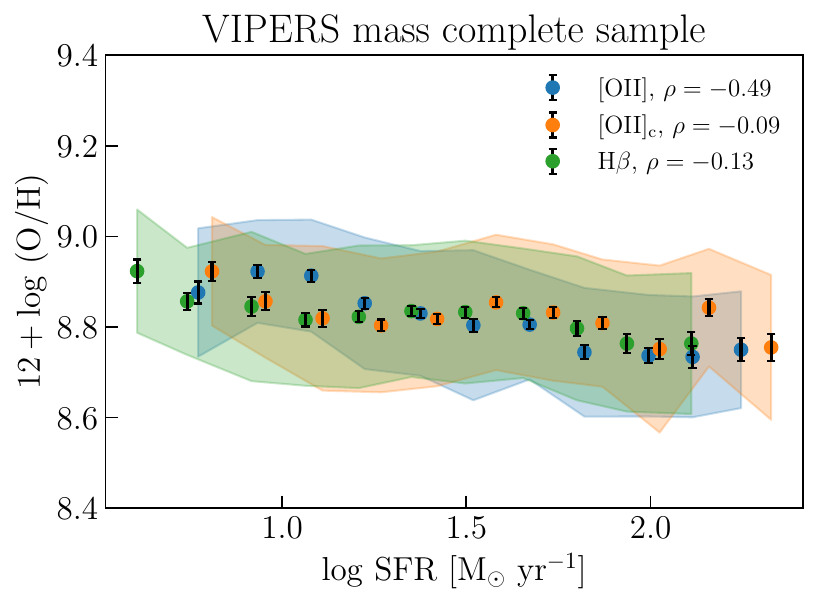}}
    \resizebox{\hsize}{!}{\includegraphics{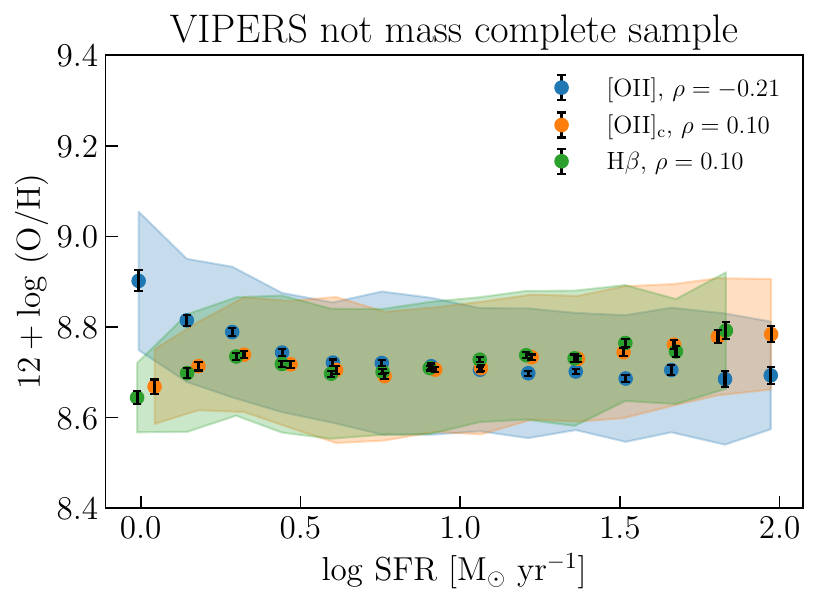}}
    \caption{Metallicity-SFR relation for different calibrations (blue: SFR derived from $\left[ \ion{O}{ii} \right]$; orange: SFR derived from $\left[ \ion{O}{ii} \right]$ with metallicity correction applied; green: SFR derived from $\mathrm{H}\beta$) for VIPERS SF sample (top panel), VIPERS mass-complete sample (mid panel), and VIPERS not mass-complete sample (bottom panel). In the legend, the correlation coefficient $\rho$ is reported for each calibration.}
    \label{fig:sfrzr_diffcal}
\end{figure}

Figure~\ref{fig:sfrzr_diffcal} shows the metallicity-SFR relation using different SFR calibrations for the VIPERS SF sample and the VIPERS mass-complete sample.
The shift at low values of SFRs between different calibrations for the VIPERS SF sample, and the positive calibration are removed once a mass-complete sample is considered.
Moreover, the agreement between different SFR calibrations improves with a mass-complete sample.
\FloatBarrier

\end{appendix}

\end{document}